\pdfoutput=1
\documentclass[11pt]{article}

\usepackage[margin=1in]{geometry}
\usepackage{amsmath,amssymb}
\usepackage{graphicx}
\usepackage{booktabs}
\usepackage{siunitx}
\usepackage{enumitem}
\usepackage[numbers,sort&compress]{natbib}
\usepackage{multibib}
\newcites{meth}{References}
\bibliographystylemeth{unsrtnat}

\usepackage{textcomp}
\usepackage{mathrsfs}
\usepackage[utf8]{inputenc}
\usepackage{subcaption}
\usepackage[hidelinks]{hyperref}

\title{Generative data assimilation highlights fronts as key regulators of ocean energy cascade}

\date{} 
\author{
Scott A. Martin$^{1}$\thanks{\texttt{smart1n@uw.edu}} \and
Georgy E. Manucharyan$^{1}$\thanks{\texttt{gmanuch@uw.edu}} \and
Patrice Klein$^{2,3}$\thanks{\texttt{pklein@caltech.edu}}
}
\date{
$^{1}$School of Oceanography, University of Washington, Seattle, USA\\
$^{2}$Environmental Science and Engineering, California Institute of Technology, Pasadena, USA\\
$^{3}$ENS, PSL Universit\'e, Ecole Polytechnique, Sorbonne Universit\'e,
CNRS, Paris, France\\
}

\begin{document}

\maketitle


\begin{abstract}
Mesoscale eddies are fundamental to the ocean circulation, yet the extent to which submesoscale motions, a few kilometers across, influence mesoscale eddy energetics through a kinetic energy cascade remains uncertain. High-resolution simulations predict that submesoscale fronts are key regulators of the cascade, transferring energy both downscale towards dissipation and upscale to sustain and shape the seasonality of mesoscale eddies. Testing these predictions has remained difficult because existing observations and state estimates cannot resolve submesoscale currents over sufficiently broad domains. Here we map the ocean's submesoscale energy cascade by combining multi-source satellite observations with a generative deep learning framework, reconstructing gap-free, kilometer-scale surface currents with physically plausible dynamics learned from simulations. Applying this to the eddy-rich Agulhas Current system, we find that submesoscales energize the mesoscale through an upscale energy cascade above 10~km, contributing to the seasonality of mesoscale eddies. Below 10 km, convergence at submesoscale fronts drives a downscale cascade towards dissipation. Both upscale and downscale pathways concentrate within fronts, where cross-scale transfer is up to an order of magnitude more efficient. Despite their limited extent, fronts account for a substantial fraction of the domain-integrated cascade, establishing them as key regulators of the cascade and targets for next-generation eddy parameterizations.
\end{abstract}


\section*{Introduction}

Global atmosphere--ocean models have highlighted the role of ocean mesoscale eddies in shaping the climate system \citep{chang2020unprecedented,seo2023ocean,wills2024resolving,simpson2025confronting}. Yet the sources and sinks that set their energy, and its variability in space and time, remain only partly quantified. While the baroclinic instability of large-scale currents is likely the primary source of mesoscale kinetic energy (KE) \citep{vallis2017atmospheric,tulloch2011scales}, recent high-resolution simulations point to an additional contributor to mesoscale KE and its seasonal variability: the fronts and eddies of the submesoscale ($\sim$1--50 km) \citep{taylor2023submesoscale} that may transfer KE upscale to the much larger mesoscale through a turbulent KE cascade \citep{sasaki2014cascade,qiu2014seasonal,schubert2020submesoscale,srinivasan2023forward}. Since the submesoscale lies far below the resolution of climate models, quantifying how much it shapes mesoscale KE is essential to establishing whether new parameterizations are needed to capture this effect.

Simulations suggest that submesoscale fronts, sharp density gradients $\mathcal{O}(10)$ km across, act as an intermediate engine helping to sustain mesoscale eddies through an upscale KE cascade. This upscale pathway is strongly seasonal: wintertime mixed-layer deepening preferentially energizes the submesoscale \citep{callies2015seasonality,taylor2023submesoscale}, and some of this KE may subsequently cascade upscale to drive a significant fraction of the seasonal cycle of mesoscale eddy energy, which often lags that at submesoscales \citep{sasaki2014cascade,qiu2014seasonal,feng2026submesoscale}. At smaller scales the cascade is expected to reverse, carrying energy downscale toward dissipation \citep{schubert2020submesoscale,srinivasan2023forward}. Some simulations further predict that both the up- and downscale branches of the cascade concentrate in frontal regions at the periphery of mesoscale eddies, where strain and convergence are strongest \citep{schubert2020submesoscale,srinivasan2023forward}.

However, evidence for this outsized role of submesoscale frontal dynamics in shaping mesoscale KE comes primarily from free-running numerical simulations. Their (sub)mesoscale energetics are sensitive to the compounding effects of limited resolution \citep{luecke2020statistical,su2018ocean}, parameterized subgrid physics \citep{large1994oceanic}, and neglected air--sea coupling \citep{strobach2022local,vivant2025ocean}, potentially producing eddy statistics that differ systematically from the real ocean. Regional \textit{in situ} campaigns provide valuable observational constraints, confirming that the upscale cascade extends into the submesoscale range \citep{balwada2022drifter,garabato2022kinetic} and that frontal convergences enhance downscale KE transfer \citep{dasaro2011enhanced,yu2024intensification}. However, their sparse spatial coverage cannot resolve the cascade's spatial distribution or quantify the contribution of small frontal regions to domain-integrated transfer.

Satellites offer a promising route for mapping the KE cascade, but their sparsity and the lack of direct surface current measurements pose major hurdles. The new Surface Water and Ocean Topography (SWOT) mission measures sea surface height (SSH) at kilometer scales across 120~km swaths with a 21-day repeat cycle \citep{fu2024surface,archer2025global}. Recent work diagnoses the cascade from SWOT SSH within individual swaths \citep{qiu2025fine,wang2025cross} by assuming geostrophic balance, whereby currents are inferred from SSH gradients \citep{vallis2017atmospheric}. However, the temporal sparsity of SWOT and the breakdown of geostrophic balance at submesoscale fronts leave much of the rapidly evolving submesoscale flow unobserved. Satellite sea surface temperature (SST) carries implicit signatures of submesoscale dynamics that recently-proposed deep learning methods exploit to infer surface currents \citep{lenain2025unprecedented}. However, clouds frequently block the observations, leaving SST-only retrievals patchy. Diagnosing the cross-scale KE fluxes that drive the cascade requires continuous, \emph{gap-free} reconstructions \citep{aluie2018coarsegrain}. Since much of the ocean is unobserved on any day, building such reconstructions requires assimilating observations across successive time steps and synergizing across variables to constrain the submesoscale over broad areas.

Existing approaches to reconstructing gap-free surface ocean dynamics face fundamental limitations at submesoscales. Physics-based DA produces physically consistent states constrained by observations, but even SWOT-assimilating systems remain limited to $\sim$70 km resolution \citep{verdy2026swot}, possibly because submesoscale dynamics in the ocean interior are poorly constrained by observations. Attention has therefore turned to deep learning: learning empirical mappings from sparse observations to gap-free surface states through regression \citep{fablet2024multimodal,martin2024deep}. However, regression-based approaches produce a single deterministic estimate, yielding overly smooth fields wherever observations do not fully constrain the submesoscale, obscuring the frontal dynamics thought to drive the cascade. Moreover, because typically no single satellite overpass covers the full domain, filling gaps requires modeling how surface fields evolve between observations rather than learning a static observation-to-state mapping.

Recently emerging generative data assimilation approaches combine the complementary strengths of DA and deep learning by learning a prior over physically plausible states, for instance from high-resolution simulations, which can then be used to generate realizations that both satisfy this prior and match available observations \citep{rozet2023score,manshausen2024generative}. Applied to surface ocean state estimation, this avoids artificial smoothing by producing ensembles of plausible small-scale realizations rather than a single deterministic prediction \citep{martin2025generative}. Relaxing the hard physical constraints of DA also allows reconstruction of the well-observed ocean surface without requiring dense interior observations. Incorporating temporal evolution during training allows observations to be assimilated across extended time horizons, increasing effective coverage \citep{rozet2023score,monkman2026spatiotemporal}.

Here we map the submesoscale KE cascade by developing a generative data assimilation framework that reconstructs gap-free, kilometer-scale surface ocean dynamics through a combination of complementary satellite observations and a generative prior learned from high-resolution simulations. We apply our framework to the Agulhas, an eddy-rich western boundary current system (Fig.~\ref{fig:study_region}) whose cross-scale interactions have been explored in simulations \citep{schubert2020submesoscale}, reconstructing the surface state over a full seasonal cycle to provide the first synoptic, observation-constrained characterization of the submesoscale KE cascade. This allows us to evaluate the contribution of submesoscale dynamics to mesoscale eddy energetics. Specifically, we focus on quantifying, for the first time, the contribution of small frontal regions to total cross-scale KE exchange.

\section*{Submesoscale-resolving surface ocean state estimates from satellite observations}\label{sec:results:skill_metrics}

Our generative data assimilation framework, \textit{GenLLC} (Methods), reconstructs kilometer-scale surface ocean states by combining sparse satellite observations with a prior learned from a high-resolution, free-running ocean simulation. The approach trains a deep learning model to generate states resembling the physics-based simulation, then guides generation with satellite observations toward the real ocean \citep{rozet2023score,martin2025generative}. Because the generated states need only preserve the structures and relationships between variables learned from the simulation rather than explicitly satisfy its governing equations, real-world observations can steer them toward configurations the free-running simulation alone may not produce while retaining physical plausibility.

This prior is implemented through a diffusion model \citep{ho2020denoising,song2020score} trained on the submesoscale-permitting global ocean simulation \textit{LLC4320} \citep{rocha2016mesoscale} to generate multivariate surface states: SSH, SST, sea surface salinity (SSS), and surface currents including their ageostrophic component. Guiding generation with submesoscale-resolving satellite observations of SSH and SST (Methods) constrains which structures appear and where. Two features of the framework are critical. First, because the prior captures temporal evolution (Methods), successive satellite passes can be assimilated jointly, building broad coverage from non-concurrent observations even as the underlying eddy field evolves. Second, because the model learns relationships among variables, observations of one field constrain the others where unobserved. For example, prior work applying generative data assimilation at mesoscales showed that SST observations constrain frontal structures in SSS and surface currents \citep{martin2025generative}. GenLLC extends this framework into the submesoscale regime, producing estimates that are observationally constrained near SWOT SSH or satellite SST observations while relying on the learned prior to evolve fields between observations and infer unobserved variables.

\begin{figure}[t]
\includegraphics[width=\textwidth]{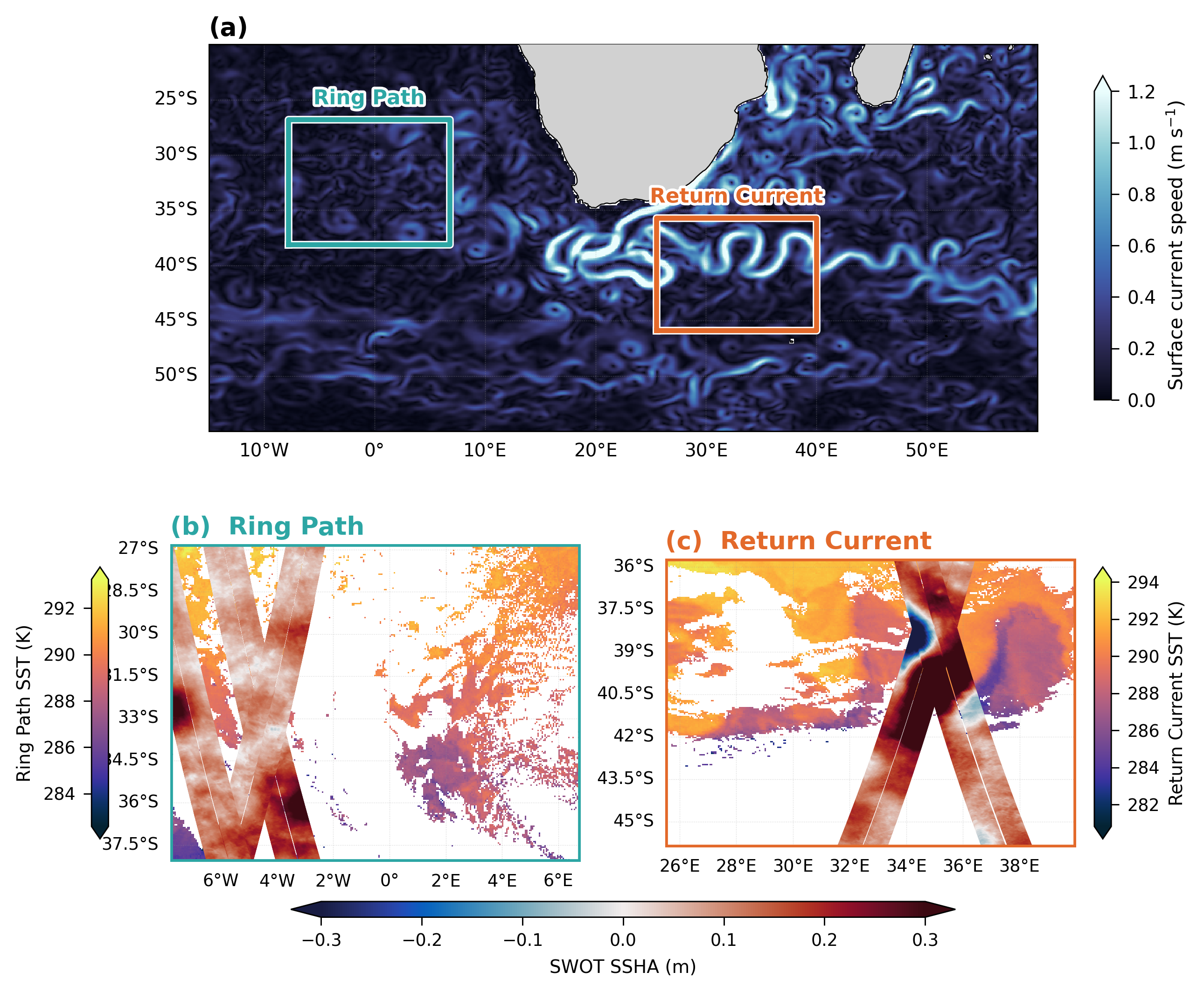}
\caption{Submesoscale-resolving, multi-modal satellite observations of Agulhas Current system. (a) Surface geostrophic currents from gridded altimetry (NeurOST) show the oceanographic context of the two study regions considered in this study. Satellite observations of SSH and SST from SWOT and SEVIRI respectively in the (b) Ring Path and (c) Return Current regions.}\label{fig:study_region}
\end{figure}

We validate GenLLC against withheld satellite SSH and SST observations and with an observing system simulation experiment (OSSE), where synthetic observations sampled from withheld LLC4320 fields enable comparison with the full ground truth, including unobserved surface currents (Methods). We apply GenLLC to the Agulhas system, spanning the Indian, Atlantic, and Southern Ocean basins (Methods), where energetic (sub)mesoscale turbulence and prior studies of submesoscale--mesoscale interactions \citep{schubert2020submesoscale} make it a natural testbed for probing the cascade. We compare two dynamically distinct sub-regions of the Agulhas system: the energetic Return Current, bisected by a meandering jet, and the more quiescent Ring Path, traversed by mesoscale rings shed from the Agulhas Retroflection (Fig.~\ref{fig:study_region}; Methods). Using satellite observations spanning the 2024 seasonal cycle, we reconstruct surface ocean states during periods with the most complete SWOT and high-resolution SST coverage, then composite the resulting reconstructions to characterize the cascade across seasons and regions (Methods).

\begin{figure}[t]
\includegraphics[width=\textwidth]{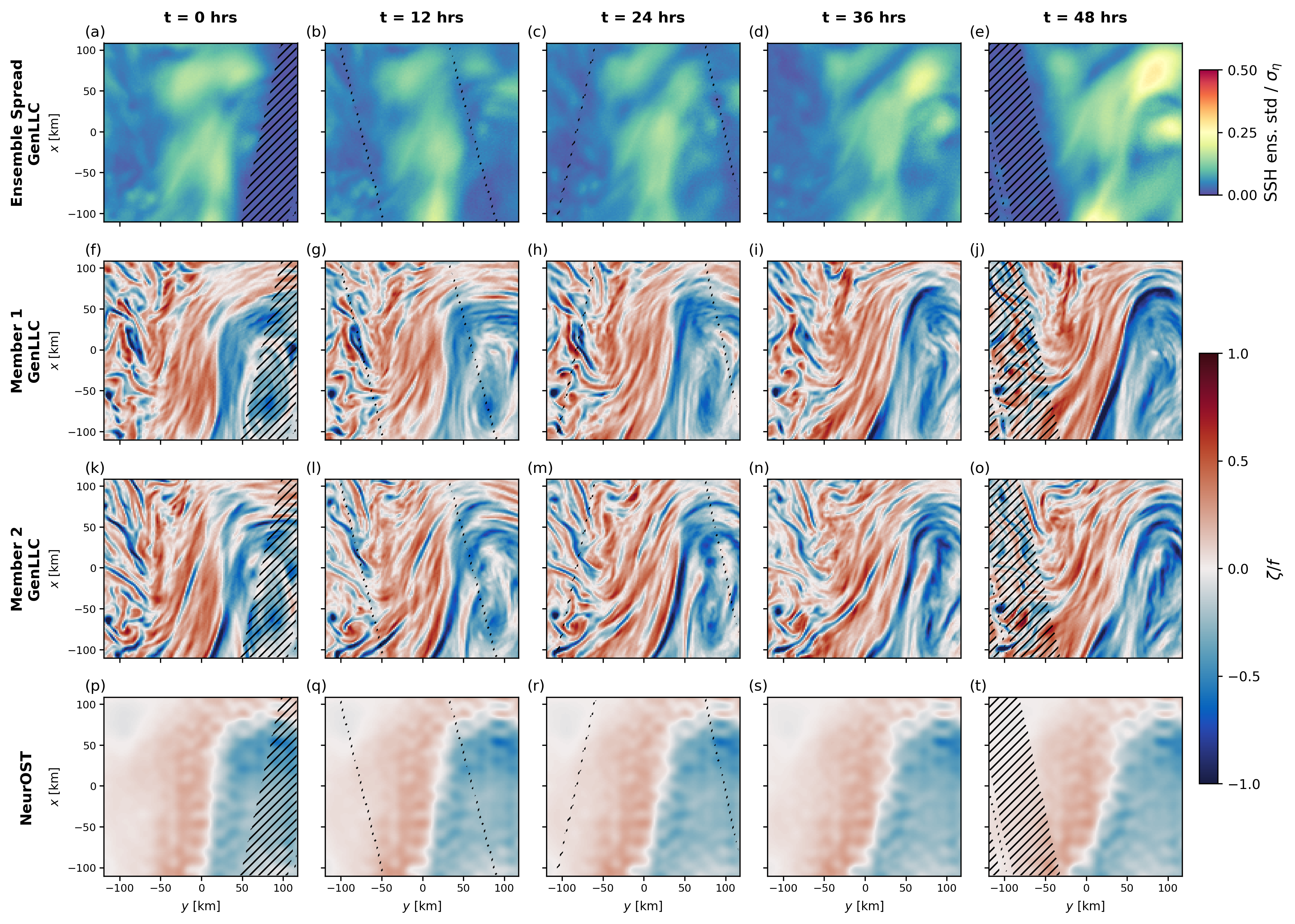}
\caption{Generative data assimilation creates an ensemble of physically plausible state estimates. (a-e) GenLLC ensemble standard deviation for predicted SSH field normalized by the standard deviation of LLC4320 SSH, with assimilated altimeter observations overlaid in hatching and time progressing from left to right. (f-j) and (k-o) reconstructed vorticity from two different GenLLC ensemble members. (p-t) Reconstructed vorticity from NeurOST gridded altimetry.}\label{fig:time_series}
\end{figure}

GenLLC reconstructs surface currents in the energetic Return Current with submesoscale-resolving fidelity (Fig.~\ref{fig:time_series}), capturing eddies and fronts as small as 5--10 km with vorticity reaching $\mathcal{O}(1)$ Rossby number (Methods), indicative of strongly ageostrophic dynamics. These structures are absent from geostrophic currents derived from gridded altimetry and remain unresolved even in NeurOST \citep{martin2024deep}, a state-of-the-art data-driven gridded SSH product (panels p--t). GenLLC ensemble members converge at submesoscales where SWOT provides high-resolution SSH observations, with close correspondence between members down to the smallest resolved scales, while providing distinct plausible realizations in observational gaps where the ensemble spread grows (panels a--o). Unlike deterministic approaches \citep{martin2024deep,fablet2024multimodal} that smooth unresolved scales, the generative approach preserves realistic submesoscale variance throughout the domain, with uncertainty reflected in the ensemble spread. The mesoscale state remains consistent across ensemble members through assimilation of coarse satellite products at scales where they are reliable (panels p--t; Methods). Crucially, reconstructed power spectra differ substantially from the LLC4320 training reference, demonstrating that the estimates are observation-constrained rather than simply reproducing the simulation statistics (Extended Data Fig.~\ref{fig:extended:real_metrics}).

Considering the full surface ocean state, GenLLC produces coherent multivariate reconstructions across scales, with co-located signatures of submesoscale eddies and fronts across all variables (Extended Data Fig.~\ref{fig:extended:real_full_snapshot}). Notably, it reconstructs submesoscale vorticity (Methods) despite assimilating no direct current observations, exploiting learned relationships between currents and the well-observed SSH and SST fields. 
Since submesoscale-resolving surface current observations are unavailable, we evaluate satellite-observable fields against withheld SSH and SST observations (Extended Data Fig.~\ref{fig:extended:real_metrics}) and assess vorticity in the OSSE, where LLC4320 provides known ground truth (Extended Data Figs.~\ref{fig:extended:osse_full_snapshot} \& \ref{fig:extended:osse_metrics}). Across both settings, GenLLC improves on interpolated satellite products in accuracy, spectral content, and calibrated uncertainty (Supplementary Information), and the OSSE demonstrates that it can recover submesoscale vorticity structures with high fidelity (Extended Data Fig.~\ref{fig:extended:osse_full_snapshot}).
These results establish GenLLC as a state estimate that captures submesoscale variability more faithfully than either gridded satellite products, which smooth it away, or free-running simulations, which are unconstrained by observations.

\section*{A seasonal submesoscale cascade energizing the mesoscale, observed from space}

\begin{figure}[t]
\includegraphics[width=\textwidth]{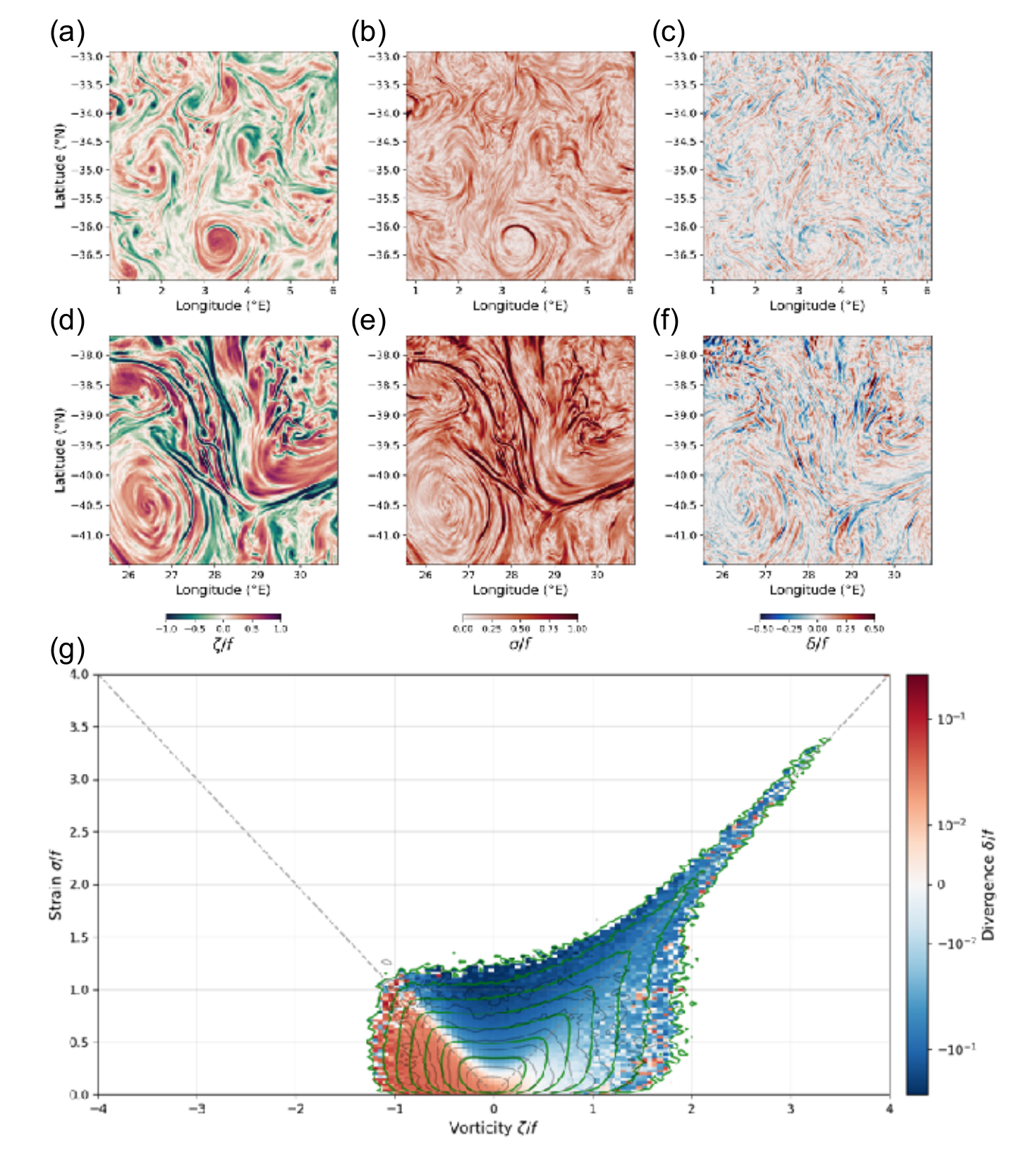}
\caption{Submesoscale surface dynamics inferred from satellite observations. GenLLC reconstructions of (a) vorticity, (b) strain, and (c) divergence in the Ring Path region, and same for the Return Current region (d-f). (g) Conditional average of GenLLC divergence conditioned on vorticity and strain. Green contours show levels of the vorticity-strain joint PDF from GenLLC, while black contours are the same from NeurOST geostrophic surface currents.}\label{fig:dynamics_main}
\end{figure}

The cross-scale energy transfers that constitute the cascade are set by the gradients of the surface currents, namely strain ($\sigma$), vorticity ($\zeta$), and divergence ($\delta$) (Methods), including their ageostrophic components \citep{aluie2018coarsegrain,srinivasan2023forward}. GenLLC recovers these gap-free gradients with a fidelity not previously achievable from space. The fields reach up to $3f$ in strain and vorticity and $0.5f$ in divergence at scales of $\mathcal{O}(10)$~km, where $f$ is the Coriolis frequency (Fig.~\ref{fig:dynamics_main}a--f), underscoring their strongly ageostrophic, submesoscale character \citep{taylor2023submesoscale} (Methods). The strong submesoscale velocities are supported by independent surface drifters, whose heavy-tailed velocity distribution GenLLC reproduces more accurately than NeurOST, albeit with a marginally higher point-wise error (Extended Data Fig.~\ref{fig:extended:drifter_validation}). Such intense submesoscale dynamics over broad domains have until now been accessible only through high-resolution simulation \citep{balwada2021vertical}.

Ageostrophic dynamics concentrate at submesoscale fronts. Conditioning divergence on vorticity and strain (Fig.~\ref{fig:dynamics_main}g) shows that the strongest convergent motions occur in strain-dominated regions ($\sigma > |\zeta|$), consistent with intense frontogenesis \citep{hoskins1982frontogenesis,taylor2023submesoscale}. The expected polarity of the frontogenetic secondary circulation, convergence where vorticity is anticyclonic and divergence where it is cyclonic \citep{hoskins1982frontogenesis,thomas2008submesoscale}, emerges below the $\sigma = |\zeta|$ line, as in high-resolution simulations \citep{balwada2021vertical}. The OSSE suggests these statistics reflect an observation-constrained view of submesoscale dynamics rather than merely a reproduction of the LLC4320 training data statistics, since GenLLC recovers the precise spatial organization of these dynamics compared against withheld LLC4320 fields (Extended Data Fig.~\ref{fig:extended:osse_vort_strain_ring_path}). This is especially notable for divergence, the hardest field to constrain: a recent SST-only inversion \citep{lenain2025unprecedented} recovered noisier divergence, with weaker positive divergence at anticyclonic fronts than obtained here. Recovering this spatial organization of surface current gradients is essential for an observation-constrained diagnosis of the KE cascade they drive \citep{capet2008mesoscale,schubert2020submesoscale,srinivasan2023forward}.

Diagnosing the cross-scale KE flux from the reconstructed currents provides the first synoptic, observation-constrained view of the submesoscale cascade (Fig.~\ref{fig:real_cascade}a). In the quiescent Ring Path, GenLLC reveals a strong upscale cascade extending from scales of $\mathcal{O}(10)$~km to the mesoscale and a downscale transfer from $\mathcal{O}(10)$~km to smaller scales. The upscale cascade intensifies markedly in winter and spring, peaking at $-3.5$~$\mu$Wm$^{-3}$, 40\% stronger than in summer and autumn. This signal is invisible to prior gridded altimetry: NeurOST resolves only mesoscale variability above $\sim$100~km and yields a flux indistinguishable from zero across the submesoscale in every season. In the OSSE, GenLLC recovers the same strength, sign, and seasonality against known LLC4320 ground truth from synthetic observations, albeit underestimating the peak upscale flux by up to 40\%, increasing confidence in the GenLLC's ability to capture the cascade despite sparse satellite sampling (Extended Data Fig.~\ref{fig:extended:osse_cascade}). Were the diagnosed cascade merely the prior painting LLC4320-like structure onto coarse observations, withholding the high-resolution SWOT and SST would leave it unchanged; instead it weakens at submesoscales and strengthens at mesoscales (Extended Data Fig.~\ref{fig:extended:high_res_obs}), showing it is genuinely constrained by the submesoscale-resolving observations.

The seasonal intensification of the submesoscale upscale cascade appears to be a major driver of the seasonal cycle of mesoscale KE. Mesoscale KE in the Ring Path typically peaks in summer \citep{steinberg2022seasonality}, and in the year reconstructed here it lags the wintertime cascade, with KE at scales above 80~km rising from $10$~Jm$^{-3}$ in early winter (June--July) to $15$~Jm$^{-3}$ by summer (November--January) (Fig.~\ref{fig:real_cascade}b). Integrating the wintertime flux over the winter--spring transition yields a mesoscale KE gain of $\mathcal{O}(10)$~Jm$^{-3}$, suggesting that the cascade is a significant contributor to summertime mesoscale energization \citep{sasaki2014cascade}. A full KE budget is not possible from surface fields alone, however, and part of this submesoscale-to-mesoscale transfer may propagate into the interior rather than emerging at the surface.

In the more energetic Return Current, cross-scale transfers are an order of magnitude larger, peaking at $-30$~$\mu$Wm$^{-3}$ with, again, a wintertime intensification of the upscale cascade (Extended Data Fig.~\ref{fig:extended:cascade_return_current}). Yet, unlike the Ring Path, multi-year altimetry shows no summertime mesoscale-KE peak here \citep{steinberg2022seasonality}. In this western boundary current regime, the upscale cascade competes with other drivers of mesoscale KE, including instability of the Return Current jet, which sheds mesoscale eddies, and upstream variability from the Agulhas Current. This contrast indicates that although submesoscales remain an active source of mesoscale KE even in the most energetic western boundary currents, their control over its variability depends on the background regime and the strength of competing larger-scale KE sources.

\section*{Frontal regions as key regulators of the bidirectional cascade}

Some simulations suggest that submesoscale fronts, at the peripheries of mesoscale eddies, are the dominant sites of cross-scale KE transfer: frontogenesis drives the downscale cascade there, while the elevated strain drives a corresponding upscale cascade at larger scales \citep{schubert2020submesoscale,srinivasan2023forward}. By recovering gap-free, submesoscale-resolving dynamics, our observation-constrained reconstructions let us test this, resolving where the cascade localizes and quantifying how much of the domain-integrated transfer the frontal regions carry. 

At scales of 5~km, where the net cascade is downscale, snapshots of the cross-scale KE flux, $\Pi_{5\,\mathrm{km}}$, reveal strong, episodic fluxes aligned with fronts around mesoscale eddies (panel~a). The conditional mean of $\Pi_{5\,\mathrm{km}}$ varies inversely with divergence (panel~b), convergent regions transferring energy downscale and divergent regions upscale, as anticipated from simulations and theory \citep{schubert2020submesoscale,srinivasan2023forward}. The downscale flux concentrates sharply in convergence zones: regions with $\delta < -0.2f$ occupy just 1.7\% of the area yet carry 28\% of the gross downscale flux (panel~c), a transfer 16-fold more efficient per unit area than the domain average. It is likewise elevated in high-strain regions (Extended Data Fig.~\ref{fig:extended:pi_v_div}a,b), since strain and convergence are co-located signatures of active frontogenesis. Frontal regions are therefore the dominant sites of downscale KE transfer, driving the cascade despite their limited extent.

\begin{figure}[t]
\includegraphics[width=\textwidth]{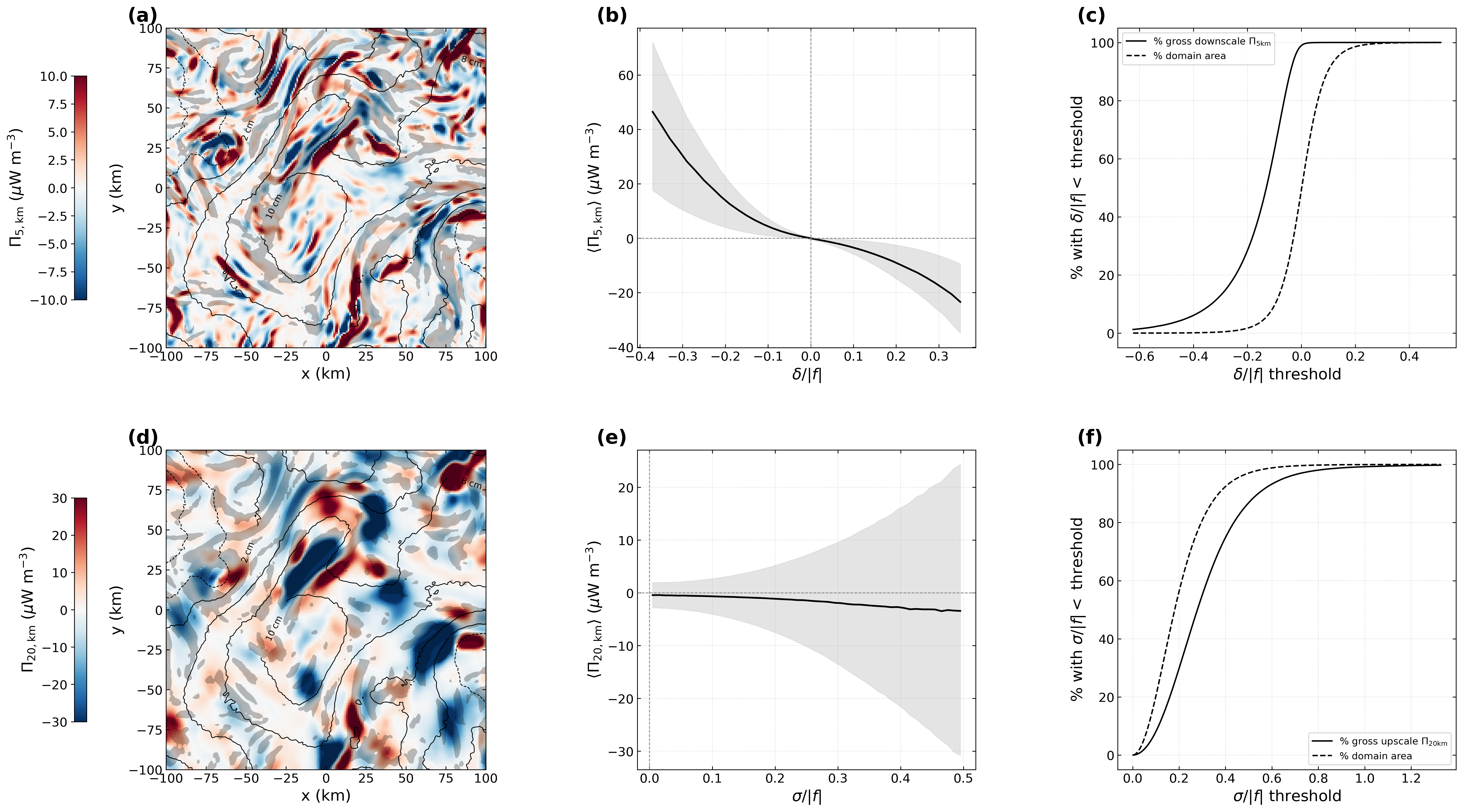}
\caption{Frontal regions around mesoscale eddies are critical in driving both the downscale and upscale cascades. (a) A snapshot of the cross-scale kinetic energy flux at 5 km with shading indicating regions with density gradient magnitude above $10^{-7}s^{-2}$. (b) Conditional mean of 5 km cross-scale kinetic energy flux against divergence. (c) Area coverage of regions with divergence below threshold (dashed) compared to the contribution of those regions to the domain-integrated gross downscale cascade at 5 km (solid). (d) A snapshot of the cross-scale kinetic energy flux at 20 km with shading indicating regions with strain rate above $0.25f$. (e) Conditional mean of 20 km cross-scale kinetic energy flux against strain. (f) Area coverage of regions with strain below threshold (dashed) compared to the contribution of those regions to the domain-integrated gross upscale cascade at 20 km (solid). SSH contours are overlaid in black in panels (a) and (d). Shading in panels (b) and (e) indicates the 16th and 84th perentiles, corresponding to the one standard deviation range for a normal distribution.}\label{fig:downscale_fronts}
\end{figure}

The upscale limb of the KE cascade is organized by these same frontal regions. At scales of 20~km, where the net cascade is upscale, the cross-scale flux, $\Pi_{20\,\mathrm{km}}$, is strongest in the high-strain regions surrounding mesoscale eddies (Fig.~\ref{fig:downscale_fronts}~d), and is again sharply localized: regions with $\sigma > 0.4f$ drive 25\% of the gross upscale flux while occupying only 7.5\% of the area (panel~f), a three-fold gain in efficiency over the domain average. Unlike at 5~km, the conditional mean trends upscale (negative $\Pi_{20\,\mathrm{km}}$) with strain (panel~e), suggesting high-strain regions route energy both up- and downscale depending on the scale considered. $\Pi_{20\,\mathrm{km}}$ still varies inversely with divergence, as at 5~km, but with far more scatter about the trend (Extended Data Fig.~\ref{fig:extended:pi_v_div}c,d). Neither strain nor divergence is a sharp local proxy for the upscale flux at 20~km, indicating it is organized at the scale of the background strain rather than set pointwise. Both limbs of the bidirectional cascade thus concentrate at the peripheries of mesoscale eddies, establishing these high-strain frontal regions, from observation-constrained state estimates, as the key regulators of cross-scale energy exchange.

\section*{Discussion}\label{sec:discussion}

These results provide the first synoptic observational evidence that submesoscale frontal dynamics transfer a significant amount of KE upscale to the mesoscale, helping to sustain mesoscale eddies. A seasonally intensified upscale cascade extends from the mesoscale down to $\mathcal{O}(10)$~km, while frontogenetic convergence simultaneously routes energy downscale toward dissipation. These pathways were predicted by high-resolution simulations \citep{sasaki2014cascade,schubert2020submesoscale,srinivasan2023forward} and glimpsed locally by \textit{in situ} campaigns \citep{balwada2022drifter,garabato2022kinetic,yu2024intensification}, but in this study are resolved over broad regions and a full seasonal cycle constrained by satellite observations, allowing us to map the submesoscale KE cascade. Crucially, by recovering the ageostrophic circulation that geostrophic altimetry cannot, our reconstructions allow us to quantify how much of the total cross-scale transfer is driven by frontal dynamics. Both limbs of the cascade are concentrated in the narrow, strained frontal regions at the peripheries of mesoscale eddies, with cross-scale KE transfers up to an order of magnitude more efficient in frontal regions. This cascade operates at scales far below those resolved by global climate models, and established eddy parameterizations represent neither the upscale transfer of energy by unresolved eddies nor its concentration in high-strain frontal regions \citep{gent1990gm_param,fox2008parameterization}. These findings further rationalize the emergence of new eddy parameterizations that take into account frontal regions \citep{zhang2023parameterizing,bodner2025data}.

While GenLLC resolves submesoscale dynamics well beyond the reach of existing satellite products, several limitations temper its interpretation. The principal one is its reliance on simulation training data from LLC4320 for the diffusion prior. Although our results demonstrate that high-resolution observations genuinely constrain the output, structural biases of the underlying physical model --- for instance an imperfect representation of symmetric instability, eddy--wave interactions, or summertime mixed-layer instability --- may persist in the learned prior and state estimates, since, as with any DA product, the estimate leverages the prior in the absence of observations. A second limitation is that SWOT SSH contains contamination from unbalanced internal waves aliased by its long repeat period \citep{gaultier2016challenge,klein2019scale_interactions}. We mitigated this by training the prior on fields with internal tides filtered out (see Methods) and by working in a region where balanced motions are expected to dominate the SSH variance at the relevant scales \citep{torres2018partitioning}. Indeed, OSSE experiments quantifying the effect of assimilating wave-contaminated SSH showed only minor degradation (Supplementary Information). Nonetheless, application in regions with energetic internal tides would benefit from emerging methods to filter these signals from observations \citep{lguensat2020filtering,gao2024deep}.
 
\begin{figure}[t]
    \includegraphics[width=\textwidth]{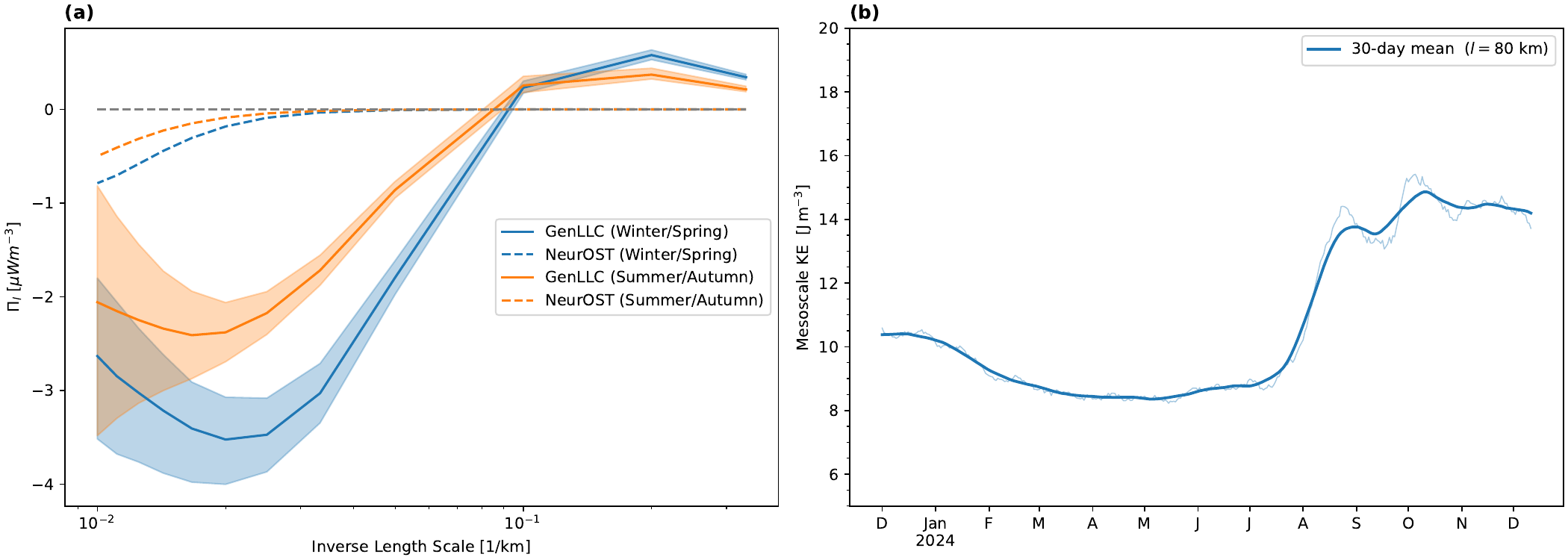}
    \caption{KE cascade estimated from real world satellite observations using GenLLC and NeurOST in the Ring Path region. (a) Cross-scale KE flux, $\Pi_l$, from GenLLC (solid) and NeurOST (dashed) as a function of coarse-graining scale, $l$. Results are split by season with winter/spring (blue) referring to June through November and summer/autumn (orange) to December through May. Shaded areas indicate the 95\% confidence interval in the GenLLC fluxes by bootstrapping over the windows in each seasonal composite. (b) Time-series of mesoscale KE (scales above 80 km) from NeurOST altimetry (dashed) along with its 30-day running mean (solid).}\label{fig:real_cascade}
\end{figure}

The comparison between our observation-constrained estimates and the LLC4320 simulation itself is instructive, though it warrants care. The cascade diagnosed from satellite observations is up to 70\% stronger than in LLC4320 in the Agulhas Ring Path (Figure \ref{fig:real_cascade} \& Extended Data Fig. \ref{fig:extended:osse_cascade}). Our analysis spans a single seasonal cycle, so part of this difference may reflect interannual variability; however, a systematic under-estimation of the upscale cascade in LLC4320 is plausible, since the model's $1/48^{\circ}$ grid only partially resolves submesoscale dynamics \citep{luecke2020statistical,archer2025global} and its fourteen-month integration may be too short for submesoscale-mesoscale interactions to equilibrate \citep{sasaki2014cascade,qiu2014seasonal}. Consistent with this interpretation, our estimates are in close alignment with a higher-resolution regional model integrated over multiple annual cycles throughout the range of scales probed here \citep{schubert2020submesoscale}. 

Beyond improving our understanding of the ocean's energy pathways, GenLLC's detailed characterization of submesoscale frontal dynamics from space enables several further directions. The divergence field, for instance, provides access to submesoscale vertical velocities \citep{torres2025submesoscale}. Vertical velocities at submesoscale fronts drive associated heat fluxes large enough to be of significance for the broader climate system \citep{su2018ocean,su2020high,siegelman2020enhanced}, and monitoring these fluxes from space would be a major step beyond the sparse campaigns available to date. Co-locating these reconstructions with satellite ocean color could reveal how submesoscale fronts structure marine ecosystems \citep{levy2018role}, while pairing them with satellite winds could test the submesoscale's emerging role in air--sea coupling, atmospheric convection, and storm development \citep{strobach2022local,vivant2025ocean,kaouah2025submesoscale}. More broadly, these findings establish generative data assimilation of sparse but complementary satellite observations as a scalable path to routine, kilometer-scale monitoring of surface ocean dynamics across the global ocean.

\bibliography{sn-bibliography}

\begin{thebibliography}{36}
\providecommand{\natexlab}[1]{#1}
\providecommand{\url}[1]{\texttt{#1}}
\expandafter\ifx\csname urlstyle\endcsname\relax
  \providecommand{\doi}[1]{doi: #1}\else
  \providecommand{\doi}{doi: \begingroup \urlstyle{rm}\Url}\fi

\bibitem[Rozet and Louppe(2023)]{rozet2023score}
Fran{\c{c}}ois Rozet and Gilles Louppe.
\newblock Score-based data assimilation.
\newblock \emph{Adv. Neural Inf. Process. Syst.}, 36:\penalty0 40521--40541,
  2023.

\bibitem[Ho et~al.(2020)Ho, Jain, and Abbeel]{ho2020denoising}
Jonathan Ho, Ajay Jain, and Pieter Abbeel.
\newblock Denoising diffusion probabilistic models.
\newblock \emph{Advances in neural information processing systems},
  33:\penalty0 6840--6851, 2020.

\bibitem[Song et~al.(2020)Song, Sohl-Dickstein, Kingma, Kumar, Ermon, and
  Poole]{song2020score}
Yang Song, Jascha Sohl-Dickstein, Diederik~P Kingma, Abhishek Kumar, Stefano
  Ermon, and Ben Poole.
\newblock Score-based generative modeling through stochastic differential
  equations.
\newblock \emph{arXiv preprint arXiv:2011.13456}, 2020.

\bibitem[Karras et~al.(2022)Karras, Aittala, Aila, and
  Laine]{karras2022elucidating}
Tero Karras, Miika Aittala, Timo Aila, and Samuli Laine.
\newblock Elucidating the design space of diffusion-based generative models.
\newblock \emph{Advances in neural information processing systems},
  35:\penalty0 26565--26577, 2022.

\bibitem[Manshausen et~al.(2025)Manshausen, Cohen, Harrington, Pathak,
  Pritchard, Garg, Mardani, Kashinath, Byrne, and
  Brenowitz]{manshausen2024generative}
Peter Manshausen, Yair Cohen, Peter Harrington, Jaideep Pathak, Mike Pritchard,
  Piyush Garg, Morteza Mardani, Karthik Kashinath, Simon Byrne, and Noah
  Brenowitz.
\newblock Generative data assimilation of sparse weather station observations
  at kilometer scales.
\newblock \emph{J. Adv. Model. Earth Syst.}, 17\penalty0 (10):\penalty0
  e2024MS004505, 2025.

\bibitem[Martin et~al.(2025)Martin, Manucharyan, and
  Klein]{martin2025generative}
Scott~A Martin, Georgy~E Manucharyan, and Patrice Klein.
\newblock Generative data assimilation for surface ocean state estimation from
  multi-modal satellite observations.
\newblock \emph{J. Adv. Model. Earth Syst.}, 17\penalty0 (8):\penalty0
  e2025MS005063, 2025.

\bibitem[Ho et~al.(2022)Ho, Salimans, Gritsenko, Chan, Norouzi, and
  Fleet]{ho2022video}
Jonathan Ho, Tim Salimans, Alexey Gritsenko, William Chan, Mohammad Norouzi,
  and David~J Fleet.
\newblock Video diffusion models.
\newblock \emph{Advances in neural information processing systems},
  35:\penalty0 8633--8646, 2022.

\bibitem[Brenowitz et~al.(2025)Brenowitz, Ge, Subramaniam, Manshausen, Gupta,
  Hall, Mardani, Vahdat, Kashinath, and Pritchard]{brenowitz2025climate}
Noah~D Brenowitz, Tao Ge, Akshay Subramaniam, Peter Manshausen, Aayush Gupta,
  David~M Hall, Morteza Mardani, Arash Vahdat, Karthik Kashinath, and Michael~S
  Pritchard.
\newblock Climate in a bottle: Towards a generative foundation model for the
  kilometer-scale global atmosphere.
\newblock \emph{arXiv preprint arXiv:2505.06474}, 2025.

\bibitem[Rocha et~al.(2016)Rocha, Chereskin, Gille, and
  Menemenlis]{rocha2016mesoscale}
Cesar~B Rocha, Teresa~K Chereskin, Sarah~T Gille, and Dimitris Menemenlis.
\newblock Mesoscale to submesoscale wavenumber spectra in drake passage.
\newblock \emph{Journal of Physical Oceanography}, 46\penalty0 (2):\penalty0
  601--620, 2016.

\bibitem[Su et~al.(2018)Su, Wang, Klein, Thompson, and Menemenlis]{su2018ocean}
Zhan Su, Jinbo Wang, Patrice Klein, Andrew~F Thompson, and Dimitris Menemenlis.
\newblock Ocean submesoscales as a key component of the global heat budget.
\newblock \emph{Nature communications}, 9\penalty0 (1):\penalty0 775, 2018.

\bibitem[Savage et~al.(2017)Savage, Arbic, Alford, Ansong, Farrar, Menemenlis,
  O'Rourke, Richman, Shriver, Voet, et~al.]{savage2017spectral}
Anna~C Savage, Brian~K Arbic, Matthew~H Alford, Joseph~K Ansong, J~Thomas
  Farrar, Dimitris Menemenlis, Amanda~K O'Rourke, James~G Richman, Jay~F
  Shriver, Gunnar Voet, et~al.
\newblock Spectral decomposition of internal gravity wave sea surface height in
  global models.
\newblock \emph{Journal of Geophysical Research: Oceans}, 122\penalty0
  (10):\penalty0 7803--7821, 2017.

\bibitem[Yu et~al.(2019)Yu, Ponte, Elipot, Menemenlis, Zaron, and
  Abernathey]{yu2019surface}
Xiaolong Yu, Aur{\'e}lien~L Ponte, Shane Elipot, Dimitris Menemenlis, Edward~D
  Zaron, and Ryan Abernathey.
\newblock Surface kinetic energy distributions in the global oceans from a
  high-resolution numerical model and surface drifter observations.
\newblock \emph{Geophysical Research Letters}, 46\penalty0 (16):\penalty0
  9757--9766, 2019.

\bibitem[Luecke et~al.(2020)Luecke, Arbic, Richman, Shriver, Alford, Ansong,
  Bassette, Buijsman, Menemenlis, Scott, et~al.]{luecke2020statistical}
Conrad~A Luecke, Brian~K Arbic, James~G Richman, Jay~F Shriver, Matthew~H
  Alford, Joseph~K Ansong, Steven~L Bassette, Maarten~C Buijsman, Dimitris
  Menemenlis, Robert~B Scott, et~al.
\newblock Statistical comparisons of temperature variance and kinetic energy in
  global ocean models and observations: Results from mesoscale to internal wave
  frequencies.
\newblock \emph{Journal of Geophysical Research: Oceans}, 125\penalty0
  (5):\penalty0 e2019JC015306, 2020.

\bibitem[Arbic et~al.(2022)Arbic, Elipot, Brasch, Menemenlis, Ponte, Shriver,
  Yu, Zaron, Alford, Buijsman, et~al.]{arbic2022frequency}
Brian~K Arbic, Shane Elipot, Jonathan~M Brasch, Dimitris Menemenlis, Aurelien~L
  Ponte, Jay~F Shriver, Xiaolong Yu, Edward~D Zaron, Matthew~H Alford,
  Maarten~C Buijsman, et~al.
\newblock Frequency dependence of near-surface oceanic kinetic energy from
  drifter observations and global high-resolution models.
\newblock \emph{arXiv preprint arXiv:2202.08877}, 2022.

\bibitem[Torres et~al.(2018)Torres, Klein, Menemenlis, Qiu, Su, Wang, Chen, and
  Fu]{torres2018partitioning}
Hector~S Torres, Patrice Klein, Dimitris Menemenlis, Bo~Qiu, Zhan Su, Jinbo
  Wang, Shuiming Chen, and Lee-Lueng Fu.
\newblock Partitioning ocean motions into balanced motions and internal gravity
  waves: A modeling study in anticipation of future space missions.
\newblock \emph{Journal of Geophysical Research: Oceans}, 123\penalty0
  (11):\penalty0 8084--8105, 2018.

\bibitem[Jones et~al.(2023)Jones, Xiao, Abernathey, and Smith]{jones2023using}
C~Spencer Jones, Qiyu Xiao, Ryan~P Abernathey, and K~Shafer Smith.
\newblock Using lagrangian filtering to remove waves from the ocean surface
  velocity field.
\newblock \emph{Journal of Advances in Modeling Earth Systems}, 15\penalty0
  (4):\penalty0 e2022MS003220, 2023.

\bibitem[Lguensat et~al.(2020)Lguensat, Fablet, Le~Sommer, Metref, Cosme,
  Ouenniche, Drumetz, and Gula]{lguensat2020filtering}
Redouane Lguensat, Ronan Fablet, Julien Le~Sommer, Sammy Metref, Emmanuel
  Cosme, Kaouther Ouenniche, Lucas Drumetz, and Jonathan Gula.
\newblock Filtering internal tides from wide-swath altimeter data using
  convolutional neural networks.
\newblock In \emph{2020 IEEE Int. Geosci. Remote Sens. Symp.}, pages
  3904--3907. IEEE, 2020.

\bibitem[Wang et~al.(2022)Wang, Grisouard, Salehipour, Nuz, Poon, and
  Ponte]{wang2022deep}
Han Wang, Nicolas Grisouard, Hesam Salehipour, Alice Nuz, Michael Poon, and
  Aur{\'e}lien~L Ponte.
\newblock A deep learning approach to extract internal tides scattered by
  geostrophic turbulence.
\newblock \emph{Geophys. Res. Lett.}, 49\penalty0 (11):\penalty0 e2022GL099400,
  2022.

\bibitem[Gao et~al.(2024)Gao, Chapron, Ma, Fablet, Febvre, Zhao, and
  Chen]{gao2024deep}
Zhanwen Gao, Bertrand Chapron, Chunyong Ma, Ronan Fablet, Quentin Febvre,
  Wenxia Zhao, and Ge~Chen.
\newblock A deep learning approach to extract balanced motions from sea surface
  height snapshot.
\newblock \emph{Geophys. Res. Lett.}, 51\penalty0 (7):\penalty0 e2023GL106623,
  2024.

\bibitem[Lyu et~al.(2024)Lyu, Wang, Pedersen, Jones, and Balwada]{lyu2024multi}
Jingwen Lyu, Yue Wang, Christian Pedersen, Spencer Jones, and Dhruv Balwada.
\newblock Multi-scale decomposition of sea surface height snapshots using
  machine learning.
\newblock \emph{arXiv preprint arXiv:2409.17354}, 2024.

\bibitem[Wang et~al.(2025)Wang, Lyu, Monkman, Jones, Pedersen, and
  Balwada]{wang2025multi}
Yue Wang, Jingwen Lyu, Tatsu Monkman, C~Spencer Jones, Christian Pedersen, and
  Dhruv Balwada.
\newblock A multi-scale probabilistic machine learning model for balanced and
  unbalanced sea surface height decomposition.
\newblock \emph{Authorea Preprints}, 2025.

\bibitem[Martin et~al.(2024)Martin, Manucharyan, and Klein]{martin2024deep}
Scott~A Martin, Georgy~E Manucharyan, and Patrice Klein.
\newblock Deep learning improves global satellite observations of ocean eddy
  dynamics.
\newblock \emph{Geophysical Research Letters}, 51\penalty0 (17):\penalty0
  e2024GL110059, 2024.

\bibitem[{OSISAF}(2017)]{seviri_sst}
{OSISAF}.
\newblock Ghrsst level 3c indian-ocean (io) sub-skin sea surface temperature
  from the spinning enhanced visible and infrared imager (seviri) on msg1 in
  gds2 format produced by osisaf, 2017.
\newblock URL
  \url{https://earthdata.nasa.gov/data/catalog/pocloud-seviri-io-sst-osisaf-l3c-v1.0-1.0}.

\bibitem[{Remote Sensing Systems}(2017)]{remss_podaac}
{Remote Sensing Systems}.
\newblock Ghrsst level 4 mw-oi global foundation sea surface temperature
  analysis version 5.0 from {REMSS} {[Dataset]}.
\newblock \url{https://doi.org/10.5067/GHMWO-4FR05}, 2017.

\bibitem[{E.U. Copernicus Marine Service Information
  (CMEMS)}(2024)]{cmems_sssl4}
{E.U. Copernicus Marine Service Information (CMEMS)}.
\newblock Multi observation global ocean sea surface salinity and sea surface
  density {[Dataset]}, 2024.
\newblock URL \url{https://doi.org/10.48670/moi-00051}.
\newblock Accessed on 09-24-2024.

\bibitem[Hoskins(1982)]{hoskins1982frontogenesis}
Brian~J Hoskins.
\newblock The mathematical theory of frontogenesis.
\newblock \emph{Annual Review of Fluid Mechanics}, 14\penalty0 (1):\penalty0
  131--151, 1982.

\bibitem[Aluie et~al.(2018)Aluie, Hecht, and Vallis]{aluie2018coarsegrain}
Hussein Aluie, Matthew Hecht, and Geoffrey~K Vallis.
\newblock Mapping the energy cascade in the {North Atlantic Ocean: T}he
  coarse-graining approach.
\newblock \emph{Journal of Physical Oceanography}, 48\penalty0 (2):\penalty0
  225--244, 2018.

\bibitem[Srinivasan et~al.(2023)Srinivasan, Barkan, and
  McWilliams]{srinivasan2023forward}
Kaushik Srinivasan, Roy Barkan, and James~C McWilliams.
\newblock A forward energy flux at submesoscales driven by frontogenesis.
\newblock \emph{Journal of Physical Oceanography}, 53\penalty0 (1):\penalty0
  287--305, 2023.

\bibitem[Vallis(2017)]{vallis2017atmospheric}
Geoffrey~K Vallis.
\newblock \emph{Atmospheric and oceanic fluid dynamics}.
\newblock Cambridge University Press, 2017.

\bibitem[Torres et~al.(2025)Torres, Wineteer, Rodriguez, Klein, Thompson,
  Perkovic-Martin, Molemaker, Hypolite, Callies, Farrar,
  et~al.]{torres2025submesoscale}
Hector~S Torres, Alexander Wineteer, Ernesto Rodriguez, Patrice Klein, Andrew~F
  Thompson, Dragana Perkovic-Martin, Jeroen Molemaker, Delphine Hypolite,
  J{\"o}rn Callies, J~Thomas Farrar, et~al.
\newblock Submesoscale eddy contribution to ocean vertical heat flux diagnosed
  from airborne observations.
\newblock \emph{Geophysical Research Letters}, 52\penalty0 (2):\penalty0
  e2024GL112278, 2025.

\bibitem[Taylor and Thompson(2023)]{taylor2023submesoscale}
John~R Taylor and Andrew~F Thompson.
\newblock Submesoscale dynamics in the upper ocean.
\newblock \emph{Annual Review of Fluid Mechanics}, 55:\penalty0 103--127, 2023.

\bibitem[Storer and Aluie(2023)]{storer2023flowsieve}
Benjamin~A Storer and Hussein Aluie.
\newblock Flow{S}ieve: A coarse-graining utility for geophysical flows on the
  sphere.
\newblock \emph{Journal of Open Source Software}, 8\penalty0 (84):\penalty0
  4277, 2023.

\bibitem[Gneiting and Raftery(2007)]{gneiting2007strictly}
Tilmann Gneiting and Adrian~E Raftery.
\newblock Strictly proper scoring rules, prediction, and estimation.
\newblock \emph{Journal of the American statistical Association}, 102\penalty0
  (477):\penalty0 359--378, 2007.

\bibitem[Hersbach(2000)]{hersbach2000decomposition}
Hans Hersbach.
\newblock Decomposition of the continuous ranked probability score for ensemble
  prediction systems.
\newblock \emph{Weather and Forecasting}, 15\penalty0 (5):\penalty0 559--570,
  2000.

\bibitem[Coadou-Chaventon et~al.(2025)Coadou-Chaventon, Swart, Novelli, and
  Speich]{coadou2025resolving}
Solange Coadou-Chaventon, Sebastiaan Swart, Guillaume Novelli, and Sabrina
  Speich.
\newblock Resolving sharper fronts of the agulhas current retroflection using
  swot altimetry.
\newblock \emph{Geophysical Research Letters}, 52\penalty0 (9):\penalty0
  e2025GL115203, 2025.

\bibitem[Lenain et~al.(2025)Lenain, Srinivasan, Barkan, and
  Pizzo]{lenain2025unprecedented}
Luc Lenain, Kaushik Srinivasan, Roy Barkan, and Nick Pizzo.
\newblock An unprecedented view of ocean currents from geostationary
  satellites.
\newblock \emph{Research Square Pre-Print}, 2025.

\end{thebibliography}

\clearpage

\clearpage

\section*{Methods}\label{sec:methods}

\subsection*{Score-based data assimilation} 

The generative data assimilation system for submesoscale-resolving surface ocean state estimation developed in this study builds on the the `score-based data assimilation' (SDA) method of \citetmeth{rozet2023score}. SDA is a method for solving inverse problems using a data-driven prior learned from simulation data, implemented in practice as an unconditional diffusion model \citepmeth{ho2020denoising,song2020score,karras2022elucidating}. We seek to infer the complete surface dynamical state, $x$, from sparse satellite observations, $y$, related through an observation operator $\mathcal{A}(x)$. The core idea is to first train a diffusion model on high-resolution simulation output to learn a prior $p(x)$ over physically consistent surface ocean states, and then use this prior within a Bayesian framework to find the state maximizing the posterior, $p(x\vert y)$.

The diffusion process in diffusion models defines a continuous transformation along a monotonically increasing `time' coordinate, $\tau$, from the data distribution at $\tau=0$, in our case the distribution of plausible surface ocean states in the simulation, to pure Gaussian noise at $\tau=T$ by progressively adding noise to the state. This process can be reversed by integrating a stochastic differential equation that depends only on the `score function', $\nabla_x\log p(x(\tau))$, meaning new samples can be drawn from the data distribution by starting from random noise and integrating backwards along $\tau$. In practice, the score function is learned by training a neural network to remove additive Gaussian noise at a wide range of noise levels \citepmeth{song2020score,karras2022elucidating}.

To perform state estimation from observations, we wish to sample from the posterior distribution rather than the prior. In SDA, to sample from the posterior distribution given observations, the posterior score, $\nabla_x\log p(x(\tau)\vert y)$, is approximated by augmenting the learned unconditional score with an approximation of the observation likelihood \citepmeth{rozet2023score} which can be evaluated given only the unconditional score network and the observation operator:
\begin{align}\label{eqn:posterior_score}
    \nabla_x\log p(x(\tau)\vert y) &= \nabla_x\log p(x(\tau)) + \nabla_x\log p(y\vert x(\tau)) \\
    &\approx \nabla_x\log p(x(\tau)) + \nabla_x\log \mathcal{N}(y\vert \mathcal{A}(\hat{x}(\tau)), \Sigma_y(\tau)),
\end{align}
where $\hat{x}(\tau)$ is the denoised state predicted by the network, and $\Sigma_y(\tau)$ is a heuristic variance that increases with $\tau$ \citepmeth{rozet2023score,manshausen2024generative,martin2025generative}.

State estimation is thus achieved by reverse-time integration from random noise, guided jointly by the learned prior and the observations. This process yields a physically plausible state consistent with $y$. Ensemble state estimates can be generated by starting from different random noise initial conditions. \citetmeth{rozet2023score}, \citetmeth{manshausen2024generative}, and \citetmeth{martin2025generative} provide full details of the approximated observation likelihood term, $\nabla_x\log p(y\vert x(\tau))$, and prior applications; below we describe the refinements beyond the method of \citetmeth{martin2025generative} made for submesoscale-resolving state estimation.

\subsection*{Learning surface ocean dynamics through video diffusion} 

In \citetmeth{martin2025generative}, we used SDA for surface ocean state estimation, training a diffusion prior to generate a state vector, $x$, where $x$ corresponds to a single time snapshot of SSH, SST, sea surface salinity (SSS), and the zonal and meridional components of surface ocean currents. While this allows learning relationships between variables, the lack of any explicit handling of time evolution means that the method draws independent samples each time-step and thus cannot generate coherent dynamical evolution where there are large gaps between observations. This becomes a key limitation for submesoscale-resolving state estimation as observations at a single time give only sparse, localized coverage. The width of the SWOT swath is 120 km, which sets too restrictive a domain size for robustly characterizing the submesoscale cascade. In addition, satellite observations of SST are missed due to the presence of clouds. By integrating observations over several consecutive time-steps, the coverage can be greatly expanded by exploiting the precessing orbit of SWOT and the time evolution of the clouds that block SST observations. To assimilate observations over multiple time-steps, a spatiotemporal prior is required to model the evolution of submesoscale eddies, fronts, and filaments between successive observations.

In this study, we thus train a diffusion prior to generate a state vector which captures a number of sequential time snapshots. This way our model can implicitly learn the {\em dynamics} of the surface ocean, with rich phenomena like eddy merging and splitting, interacting eddy dipoles, and strain-driven deformation around eddy cores that are all captured in the high-resolution simulation training data. In practice, we train a diffusion prior to generate five consecutive time-steps which we will demonstrate is sufficient to learn physically-realistic eddy evolution and to provide broad spatial coverage when observations are assimilated over the full time-series. We model the time dimension explicitly by utilizing a video diffusion architecture \citepmeth{ho2022video}. Specifically, in the neural network architecture of our diffusion model, we apply the same 2D spatial convolutional filters to each time-step independently, before applying pixel-wise temporal attention to model the time dynamics. Applying temporal attention at different depths in the backbone allows the model to capture multi-scale dynamics. This architecture is inspired by that used in the global atmosphere foundation model Climate in a Bottle \citepmeth{brenowitz2025climate}. This design allows assimilation of observations over multiple time-steps. Employing a prior that captures time evolution also provides a significantly stronger physical constraint on the reconstructed states than a purely spatial prior. Details of the neural network architecture and SDA hyperparameters used in GenLLC are in the Supplementary Information.

\subsection*{Submesoscale-permitting ocean model simulation training data} 

To provide a large training dataset of physically plausible realizations of (sub)mesoscale ocean dynamics, we train our diffusion prior on surface data from the NASA LLC4320 simulation, a global 1/$48^{\circ}$ MITgcm simulation \citepmeth{rocha2016mesoscale,su2018ocean}. The LLC4320 simulation is the highest-resolution global ocean simulation run with realistic forcing for a long enough period to cover a full annual cycle. More details about the configuration of LLC4320 can be found in \citetmeth{rocha2016mesoscale}. The simulation covers fourteen months of output at $1/48^{\circ}$ resolution, spun up using using increasingly high resolution simulations which were initially initialized using ECCO initial conditions to ensure a large-scale ocean state consistent with real observations. We discard the first two months and train our prior on the final full annual cycle of output to mitigate the effects of spin-up when LLC4320 is adjusting to the increase in model resolution from its low-resolution parent simulation. We focus our study on the region connecting the Atlantic and Indian ocean basins encompassing the Agulhas Retroflection, Agulhas Return Current, and extend the domain to the Indo-Atlantic sector of the Antarctica Circumpolar Current, covering $5^{\circ}$ to $40^{\circ}$ E, and from $35^{\circ}$ to $55^{\circ}$ S. We focus on this region since it features energetic (sub)mesoscale eddy turbulence and allows us to avoid training on the full range of global ocean dynamics while maintaining a large training domain to facilitate training on a diverse training dataset.


We train our diffusion prior to generate SSH, SST, SSS, and both zonal and meridional surface currents from LLC4320. The neural network processes patches of spatial size 256 by 256 pixels where the data are presented on the native LLC4320 grid, which in this region has a typical spacing of 1.5-2 km. We train a diffusion model to generate a series of 5 time-steps, where each time-step is a 12-hour average of the hourly simulation output. These choices were made to make the network's field of view sufficient to capture a range of (sub)mesoscale dynamics while staying within reasonable computational bounds. For SST and SSS, we remove the patch mean so that the network learns to generate SST and SSS anomalies, while for SSH we subtract a time-varying linear plane fit to remove both the large-scale background and the effects of barotropic tides and the inverse barometer effect which both impact SSH in LLC4320 but are filtered out from real world altimeter observations. For surface currents, we take the current at 15 m depth to reduce the impacts of surface winds and further pre-process by subtracting the patch mean to remove large-scale signals which can be easily inferred in the real world setting from existing methods and data products. We generate training examples by sub-sampling from our full training domain and excluding any examples containing land or sea ice in any pixel.

Since LLC4320 is known to have overly energetic internal tides which project strongly onto SSH and surface currents \citepmeth{savage2017spectral,yu2019surface,luecke2020statistical,arbic2022frequency}, we apply a sub-inertial filter to the SSH and surface current fields to remove the internal tides before taking the 12-hourly average. We apply a low-pass temporal filter to the data removing signals with frequency above 1.1$f$, where $f$ is the local inertial frequency. While more sophisticated approaches have been proposed in previous work to more cleanly isolate the signature of internal tides \citepmeth{torres2018partitioning,jones2023using}, this approach is a compromise between efficacy and ease of implementation. While this step allows us to train a diffusion prior to generate ocean states with the internal tides (mostly) filtered out, when assimilating real world SWOT SSH observations, sub-inertial unbalanced wave signals will be present in the data. Developing a method to remove internal tides from SWOT observations is an active area of research \citepmeth{lguensat2020filtering,wang2022deep,gao2024deep,lyu2024multi,wang2025multi}. Contamination from unbalanced internal tides may thus be present after applying our generative data assimilation approach to real SWOT observations. We perform additional experiments to assess the impact of this on the quality of the reconstructions. The contamination of unbalanced internal tides is expected to be most significant in local summertime when the mixed layer is shallow, but in our study region \citetmeth{torres2018partitioning} found balanced motions to be the leading order contributor to SSH and surface current variance in the 10-100 km scale range throughout the year.

We create a large training dataset of patches from the study region that sample a full annual cycle of LLC4320 output. This exposes the diffusion model to a diverse range of (sub)mesoscale surface ocean dynamics. Since the model learns local dynamics and is given no input information about geographical location, the learned prior will capture a wide range of physically plausible dynamics. We expect that the diversity of states seen during training will help the model to generalize across dynamical regimes if supported by the observations during state estimation.

\subsection*{Observation operator for submesoscale surface ocean state estimation}

Building upon the method presented in \citetmeth{martin2025generative}, we design an observation operator to exploit multi-modal, multi-resolution satellite observations of SSH, SST, and SSS. For SSH, we combine three observation sources: submesoscale-resolving wide-swath SWOT observations, along-track nadir altimeter observations, and coarse gridded SSH estimates from NeurOST \citepmeth{martin2024deep}. For the sparse SWOT and nadir altimeter observations, the observation operator sub-samples the predicted SSH at the grid points observed by satellites, while to assimilate the coarse NeurOST SSH we first coarse-grain the prediction with a filter chosen to reflect the effective resolution of NeurOST. Assimilating coarse satellite products in this way allows us to leverage the accuracy of these products at large scales without penalizing the diffusion prior for generating variance beneath the coarse products' effective resolution \citepmeth{martin2025generative}. For SST, we leverage geostationary observations from SEVIRI \citepmeth{seviri_sst} to obtain high-resolution, cloud-occluded SST measurements with consistent sampling and instrument errors owing to being sampled by a single platform and further assimilate coarse SST estimates from objective analysis \citepmeth{remss_podaac}. Since the resolution of SEVIRI SST is $\sim0.05^{\circ}$ in our study region, we first coarse-grain the predicted SST to this resolution before masking out pixels where the SEVIRI measurements are occluded by clouds. For SSS, we assimilate only coarse satellite estimates from objective analysis \citepmeth{cmems_sssl4} as high-resolution satellite SSS observations are not available, and for surface currents we assimilate only large-scale NeurOST geostrophic currents, since satellites do not observe small-scale surface currents and the dynamics at large scales are mostly geostrophic. Details of the precise coarse-graining scales and specific satellite products are in Supplementary Information. Taken together, the observation operator allows us to exploit high-resolution but gappy observations of SSH and SST to constrain the submesoscale state. Then, the multi-variate learned prior transfers these submesoscale observational constraints to the more coarsely observed SSS and surface current fields.

\subsection*{Geostrophy and characterizing dynamics through surface current gradients}

Throughout the manuscript, we characterize surface ocean dynamics through the gradients of surface current velocities. Specifically, we consider three components of the velocity gradient tensor which together characterize the evolution and deformation of tracers in geophysical flows. First, the relative vorticity,
\begin{equation}
    \zeta = \frac{\partial v}{\partial x} - \frac{\partial u}{\partial y},
\end{equation}
where $u$ and $v$ are the zonal and meridional surface current velocities respectively, and $x$ and $y$ are zonal and meridional coordinates respectively. $\zeta$ characterizes the local rotation of the flow. Second, the strain rate,
\begin{equation}
    \sigma = \sqrt{\left(\frac{\partial u}{\partial x} - \frac{\partial v}{\partial y}\right)^2 + \left(\frac{\partial v}{\partial x} + \frac{\partial u}{\partial y}\right)^2},
\end{equation}
which characterizes the stretching of fluid elements. High strain rate is associated with enhanced frontogenesis \citepmeth{hoskins1982frontogenesis} and cross-scale KE exchange \citepmeth{aluie2018coarsegrain,srinivasan2023forward}. Finally, the divergence,
\begin{equation}
    \delta = \frac{\partial u}{\partial x} + \frac{\partial v}{\partial y},
\end{equation}
which characterizes the extent to which the velocity field is converging or diverging and is closely connected to the local vertical velocity \citepmeth{vallis2017atmospheric,torres2025submesoscale}, with $\delta < 0$ (convergence) associated with downwelling and $\delta > 0$ with upwelling.

Geostrophic balance refers to the dominant force balance at large scales in geophysical flows, where the Coriolis force balances the pressure gradient force \citepmeth{vallis2017atmospheric}. The geostrophic approximation is widely used to infer surface ocean currents from satellite observations of SSH,
\begin{equation}
(u_g, v_g) = \frac{g}{f}\left(-\frac{\partial \eta}{\partial y}, \frac{\partial \eta}{\partial x}\right),
\end{equation}
where $\eta$ is the SSH, $g$ the acceleration due to gravity, $f$ the Coriolis frequency, and $u_g$ and $v_g$ are the geostrophic components of the surface current velocity. Since the geostrophic approximation increasingly breaks down at submesoscales \citepmeth{taylor2023submesoscale} the surface currents have a non-negligible \textit{ageostrophic} component,
\begin{equation}
    (u_{ag}, v_{ag}) = (u - u_g, v - v_g),
\end{equation}
which is not retrievable from SSH alone but which we seek to capture in GenLLC by training the prior to predict the total velocities ($u$, $v$) including the ageostrophic contribution. Surface currents have a significant ageostrophic contribution when the Rossby number, $\mathcal{R}o=\zeta/f$, exceeds $\sim0.1$.

\subsection*{Coarse-graining computation of cross-scale kinetic energy fluxes}

The kinetic energy (KE) of surface ocean currents is given by
\begin{equation}
KE = \frac{1}{2}\rho_0\left(u^2 + v^2\right),
\end{equation}
where $\rho_0$ is a reference density (here taken to be 1025~kg~m$^{-3}$).

The turbulent KE cascade manifests as cross-scale KE fluxes, $\Pi_l$, which characterize the amount of kinetic energy transferred from scales larger than $l$ to smaller scales by the non-linear interaction term in the Navier-Stokes equation \citepmeth{aluie2018coarsegrain}. $\Pi_l$ is characterized by the interplay between the large-scale strain tensor, $\mathbf{\overline{S}}_l$, and the subfilter-scale stress, $\mathbf{\overline{\tau}}_l$, where the overbar denotes a coarse-graining (spatial low-pass filtering) operation at length scale $l$, through \cite{aluie2018coarsegrain}
\begin{equation}\label{pi_definition}
    \Pi_l = -\rho_0\overline{S}_{ij}\overline{\tau}_{ji},
\end{equation}
where 
\begin{align}
    \overline{S}_{ij} &= \frac{1}{2}\left(\partial_i \overline{u}_j+\partial_j \overline{u}_i\right), \\
    \overline{\tau}_{ij} &= \overline{u_i u_j} - \overline{u}_i\overline{u}_j,
\end{align}
repeated indices are summed over, and the subscript $l$ in the coarse-graining operation has been dropped when using index notation to avoid confusion between the coarse-graining length scale and a spatial index. The cascade term, $\Pi_l$, represents the energy transfer from scales larger than $l$ to smaller scales due to non-linear eddy interactions, so where $\Pi_l$ is positive (negative), energy is transferred from scales larger (smaller) than $l$ to smaller (larger) scales representing a downscale (upscale) cascade. Further scrutiny of Equation \ref{pi_definition} illustrates the sensitivity of the KE cascade to the strain rate and divergence \citepmeth{srinivasan2023forward}.

We use {\em FlowSieve}, a parallelized coarse-graining and cross-scale energy flux computation code \citepmeth{storer2023flowsieve}, to compute the strength and direction of the kinetic energy cascade which is quantified through the cross-scale energy flux, $\Pi_l$. We re-grid all data onto a regular local Cartesian grid with 1 km resolution before coarse-graining calculations using the ENU projection. For coarse gridded products, high-frequency interpolation artifacts appear after regridding to 1 km which we remove before cross-scale flux computations by applying a Gaussian low-pass filter with 3-pixel filter width. The smoothing filter used in the FlowSieve coarse-graining analysis is a smoothed top-hat, as used in previous studies \cite{storer2022global}
\begin{equation}
    G_l(\mathbf{r}) = \frac{A}{2}\left(1-\text{tanh}\left[10\left(\frac{\lvert\mathbf{r}\rvert}{l/2}-1\right)\right]\right),
\end{equation}
where $A$ is a normalization calculated numerically to ensure $G_l$ integrates to unity and $\mathbf{r}$ is the separation between the evaluation point and the center of the convolutional kernel. The coarse-grained fields, $\overline{f}_l(\mathbf{x})$, are then defined as
\begin{equation}
    \overline{f}_l(\mathbf{x}) = G_l*f,
\end{equation}
where $*$ is a two-dimensional convolution. 

\subsection*{Quantifying the contribution of frontal regions to domain-integrated KE cascade}

To assess the role of frontal dynamics in driving cross-scale KE fluxes at a given scale, we aggregate across all our reconstructions and assess how $\Pi_l$ co-varies with strain and divergence by taking the conditional mean of $\Pi_l$ in $\sigma$ or $\delta$ bins (e.g. Fig.~\ref{fig:downscale_fronts}b,e).

We further quantify the extent to which frontal regions, which typically only occupy a small fraction of the domain, contribute to the domain-integrated transfer. To do so, we separately consider grid points with downscale ($\Pi_l > 0$) and upscale ($\Pi_l < 0$) cascade, and examine how the flux accumulates as a function of a conditioning variable, $\chi$, taken to be either the strain rate, $\sigma$, or the divergence, $\delta$. For each cascade direction, we sort grid points by $\chi$ and compute the cumulative sum of $\Pi_l$ up to a given threshold value, $\chi^*$, normalized by the total flux summed over all grid points of that cascade direction:
\begin{equation}
    C_\Pi(\chi^*) = \frac{\sum_{\chi < \chi^*} \Pi_l}{\sum_{\text{all}} \Pi_l}.
\end{equation}
We compare this to the fraction of the domain area with $\chi < \chi^*$,
\begin{equation}
    C_A(\chi^*) = \frac{N(\chi < \chi^*)}{N_{\text{total}}},
\end{equation}
where $N$ denotes the number of grid points. If $C_\Pi(\chi^*) > C_A(\chi^*)$, the KE cascade is disproportionately concentrated in the low-$\chi$ portion of the domain, relative to the area it occupies; equivalently, a small fraction of the domain (by area) accounts for a large fraction of the total cascade. In the text, we sometimes quote this fraction as an increase in the efficiency of cross-scale KE transfer compared to the domain average. This corresponds to taking the ratio
\begin{equation}
    \text{Relative Efficiency} = \frac{C_\Pi(\chi^*)}{C_A(\chi^*)}.
\end{equation}

\subsection*{Experiments}

\subsubsection*{Observing system simulation experiment}\label{sec:experiments:osse}

To assess the state estimation capabilities of GenLLC in a observing system simulation experiment (OSSE) where the full ground truth is known, we evaluate it first using simulated satellite observations sampled from the simulation. Synthetic altimetry observations are generated by sampling the LLC4320 fields with observation masks taken from satellite observations for the year 2024. For SST, we apply cloud masks taken from the SEVIRI satellite observations for the year 2024 and further coarse-grain the LLC4320 SST to the resolution of the SEVIRI data product, with nominal $0.05^{\circ}$ resolution. Simulated coarse gridded satellite SSH, SST, SSS, and surface geostrophic currents are generated by coarse-graining the simulation in space and time to reflect the effective resolution of these various products (Supplementary Information). 

Throughout this study we focus on two study regions in which we explore the strength and seasonality of the submesoscale cascade which were chosen to reflect distinct background oceanographic conditions. We refer to these regions as the Agulhas Return Current ($25.5^{\circ}$ to $40.0^{\circ}$E, $35.7^{\circ}$ to $45.9^{\circ}$S) and the Agulhas Ring Path ($-7.8^{\circ}$ to $6.8^{\circ}$E, $26.8^{\circ}$ to $38.1^{\circ}$S). The Return Current is a strongly energetic region bisected by the meandering jet of the Agulhas Return Current, while the Ring Path is a relatively quiescent open ocean region with weak background currents and strong mesoscale eddies shed from the Agulhas retroflection, these Agulhas Rings periodically transit through the Ring Path region. The Return Current lies within the larger domain used to train the diffusion prior while the Ring Path is outside the training domain. This requires the prior to extrapolate to potentially distinct regional dynamics. The quantitative evaluation of the state estimates in Section \ref{sec:results:skill_metrics} for the Return Current region is conducted only on the withheld two-month spin-up phase of LLC4320 to avoid any contamination from the training data.

To ensure that the method is applied in windows where observations of SSH and SST provide strong constraints on the submesoscale surface ocean state, we sub-sample the two study regions described above in space and time to select the windows with good observational coverage of both SSH and SST. We use all SWOT and SEVIRI observations available during the year 2024 to select windows corresponding to the dimensions of $x$ by requiring a minimum of 15\% and 30\% pixels to be observed by SWOT and SEVIRI respectively. For SWOT this corresponds roughly to at least three SWOT passes through the domain during the 2.5 day sequence reconstructed by the prior, and for SEVIRI corresponds to moderate but not severe cloud cover. We select as many windows as possible subject to these constraints while also minimizing overlap between selected windows and maximizing coverage of the full annual cycle. After this procedure we are left with 191 and 204 selected windows in the Ring Path and Return Current respectively. By reconstructing small, non-uniformly distributed patches from our larger study regions some noise and bias will be introduced in the estimate of the submesoscale cascade compared to what would be found by calculating it on the full domain. In Suuplementary Information, we demonstrate that estimating the cascade in these small windows provides a consistent result with that calculated from the full domain for scales below $\mathcal{O}(100)$km.

We evaluate the point-wise accuracy of the state estimates using the Continuous Ranked Probability Score \citepmeth[CRPS;][]{gneiting2007strictly,hersbach2000decomposition}, a scoring rule that jointly assesses the accuracy and calibration of probabilistic predictions by measuring the distance between the predicted ensemble distribution and the observed value. The CRPS decomposes into the mean absolute error of the ensemble mean minus a term rewarding ensemble spread that is proportional to forecast uncertainty, penalizing overconfident predictions where ensemble spread is low but errors are large. For a deterministic prediction, CRPS reduces to the mean absolute error, providing a unified framework for comparing GenLLC ensemble predictions against deterministic products such as coarse gridded satellite products.

\subsubsection*{Application to real world satellite observations}\label{sec:experiments:real}

We apply GenLLC to real world surface ocean state estimation to provide observational constraints on the strength of the cascade. We apply the method in the same well-observed windows from both the Return Current and Ring Path regions using observations from 2024. To assimilate real observations we simply replace the simulated satellite observations generated from the simulation with SWOT and nadir altimetry and SEVIRI SST observations, and replace the simulated coarse gridded products with coarse satellite products of SSH, SST, SSS, and surface geostrophic currents.

To enable independent validation of the real world state estimates, we apply additional cloud masks to the SEVIRI SST observations and withhold SWOT observations on the second and fourth time-step of the reconstructed state. This provides independent high-resolution SSH and SST observations with which we can validate the GenLLC reconstructed SSH and SST. Since in the validation experiment we withhold a significant fraction of available observations from the assimilation, the reconstruction skill will be a lower bound on the performance of GenLLC when all available observations are assimilated. To avoid degrading the observations too drastically, we withhold the SSH and SST observations individually in two separate experiments, so we use degraded SSH with full SST and vice versa.

\section*{Data Availability}
All observation datasets used in this study are publicly available and can be accessed through the references in the text. The LLC4320 simulation data used here were extracted from the full petascale simulation output from within the secure enclave at the NASA Advanced Supercomputing (NAS) Division at Ames Research Center, but the full simulation output can be publicly accessed either through the Pangeo forge catalog \url{https://catalog.pangeo.io/browse/master/ocean/LLC4320/} for surface-only fields or the NSF-funded Poseidon project \url{https://www.poseidon-ocean.net/products/datasets/llc4320-dataset/} for all variables and depth levels. The GenLLC state estimates analyzed in this study are available through \url{https://doi.org/10.5281/zenodo.21479676}.

\section*{Code Availability}
The GenLLC code is available through a public GitHub repository \url{https://github.com/smartin98/GenLLC4320} and the code used to extract and handle SWOT observations is available through another public GitHub repository \url{https://github.com/smartin98/SwotDB}. Our implementation builds on a number of open-source codebases which we gratefully acknowledge: NVIDIA PhysicsNemo \url{https://github.com/nvidia/physicsnemo}, NVIDIA Climate in a Bottle \url{https://github.com/NVlabs/cBottle}, and Francois Rozet's Score-Based Data Assimilation \url{https://github.com/francois-rozet/sda}. The coarse-graining diagnosis of cross-scale kinetic energy fluxes in this study was carried out using FlowSieve \url{https://github.com/husseinaluie/FlowSieve}.

\section*{Author Contributions}
S.A.M. led the conceptualization, methodology, analysis, and visualization. All authors discussed and interpreted the results and contributed to the writing and editing of the manuscript.

\section*{Competing Interests}
The authors declare no competing interests.

\section*{Acknowledgments}
S.A.M. and G.E.M. were supported by NASA grant 80NSSC21K1187, and P.K. was supported by NASA grant 80NSSC24K1653. The authors gratefully acknowledge helpful discussions with Noah Brenowitz \& Tatsu Monkman. Computational resources supporting this work were provided by the NASA High-End Computing (HEC) Program through the NASA Advanced Supercomputing (NAS) Division at Ames Research Center.

\clearpage
\section*{Extended Data}

\renewcommand{\figurename}{Extended Data Fig.}
\setcounter{figure}{0}

\begin{figure}[t]
\includegraphics[width=\textwidth]{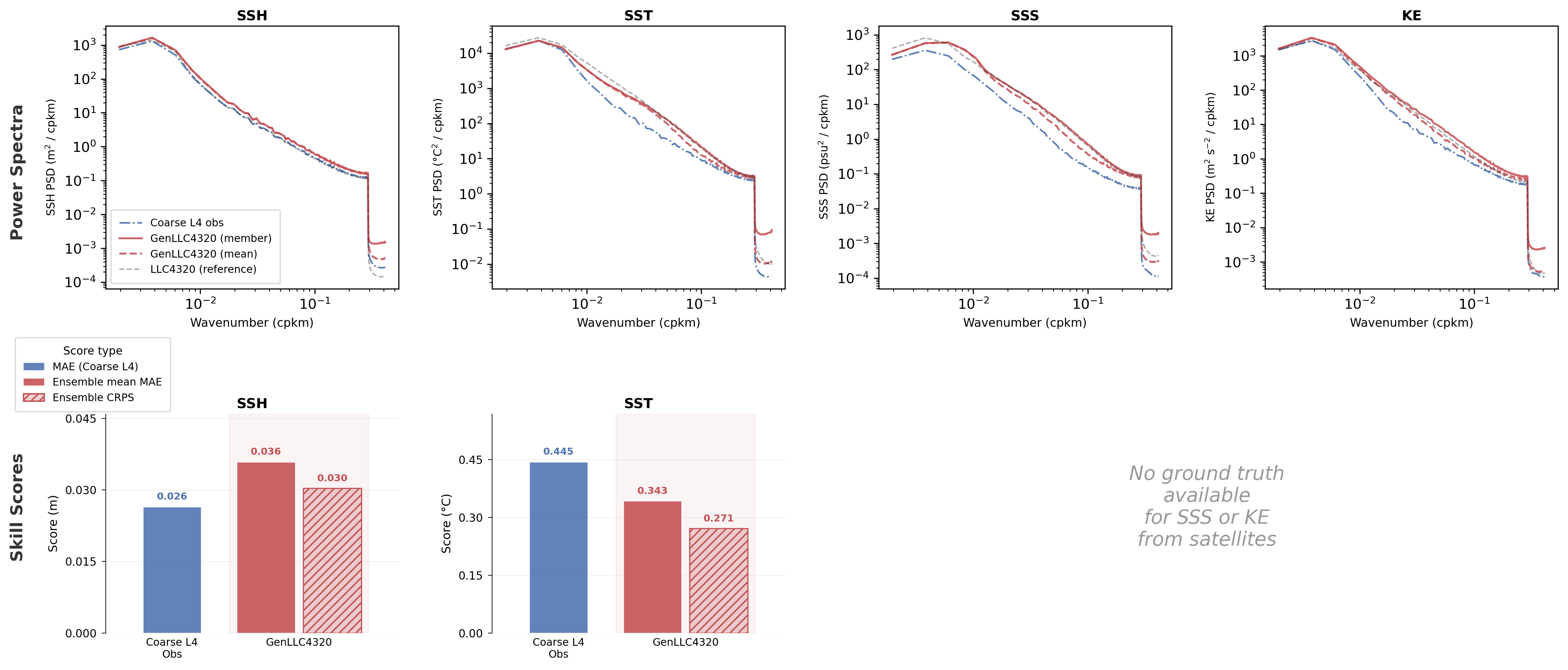}
\caption{State estimation evaluation metrics for GenLLC in the Return Current for the experiment with real world observations, aggregated over 20 observation windows each with a 10-member ensemble. Top row: comparison of the power spectra for SSH, SST, SSS, and kinetic energy for the coarse gridded satellite products (blue dot dashed), GenLLC ensemble member (red solid), GenLLC ensemble mean (red dashed), and LLC4320 is shown for reference (dashed gray) though in the real world context this is no longer the ground truth. Bottom row: point-wise accuracy comparison comparing the mean absolute error of the coarse gridded satellite products (blue) to the mean absolute error of the GenLLC ensemble mean (red solid), and the CRPS of the GenLLC ensemble members (red hatched) for SSH and SST using the withheld satellite observations. Point-wise accuracy for SSS and surface currents in the real world setting is not evaluated here due to the lack of satellite observations of these quantities.}\label{fig:extended:real_metrics}
\end{figure}

\begin{figure}[t]
\includegraphics[width=\textwidth]{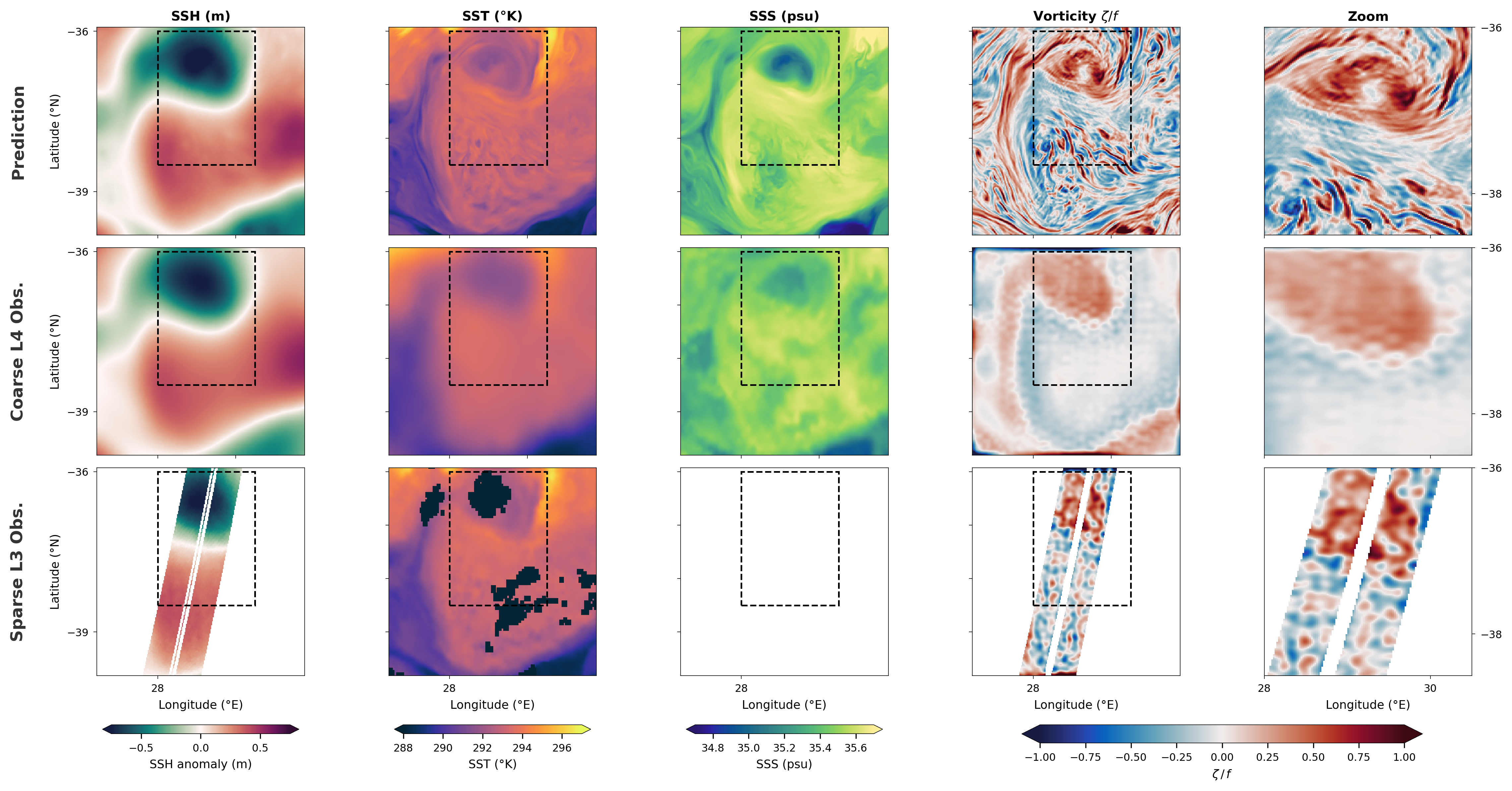}
\caption{Example GenLLC reconstruction from real satellite observations compared to operational gridded satellite products. The panels from left to right show SSH, SST, SSS, vorticity, and a zoomed in regional crop of the vorticity fields with the GenLLC state estimate in the top row, operational gridded satellite products in the middle row, and sparse high-resolution satellite observations in the bottom row. The SWOT pass shown was not assimilated in the state estimation, and the vorticity in the bottom row is computed from the SWOT SSH assuming geostrophy after applying a 6 km Gaussian low pass filter to suppress noise from the instrument.}\label{fig:extended:real_full_snapshot}
\end{figure}

\begin{figure}[t]
\includegraphics[width=\textwidth]{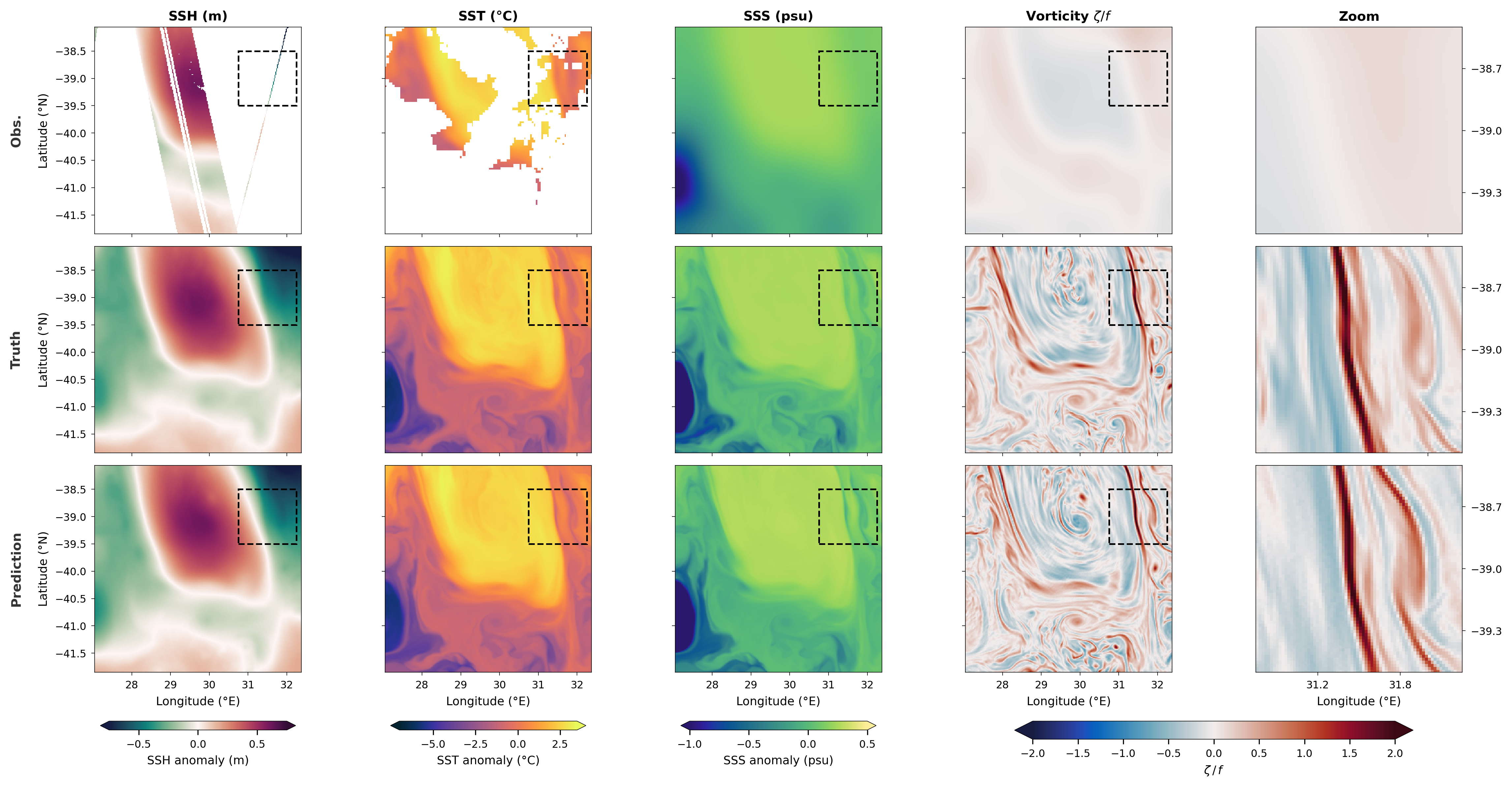}
\caption{Example GenLLC reconstruction from simulated satellite observations compared to LLC4320 ground truth. The panels from left to right show SSH, SST, SSS, vorticity, and a zoomed in regional crop of the vorticity fields with the simulated observations in the top row, the LLC4320 ground truth in the middle row, and the GenLLC state estimate in the bottom row.}\label{fig:extended:osse_full_snapshot}
\end{figure}

\begin{figure}[t]
\includegraphics[width=\textwidth]{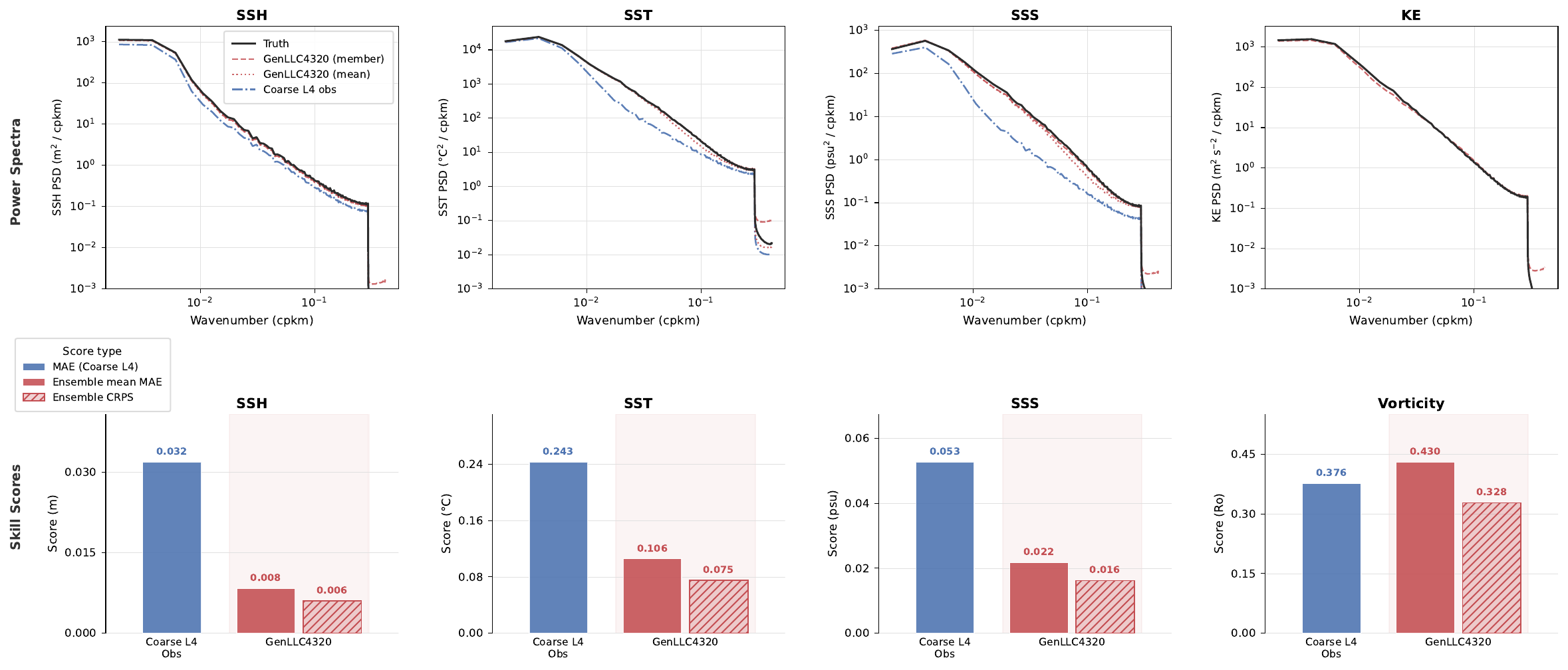}
\caption{State estimation evaluation metrics for GenLLC in the Return Current for the observing system simulation experiment, aggregated over 22 observation windows each with a 20-member ensemble. Top row: comparison of the power spectra for SSH, SST, SSS, and kinetic energy for the LLC4320 ground truth (black), GenLLC ensemble members (red dashed), GenLLC ensemble mean (red dotted), and coarse gridded satellite products (blue dot dashed). Bottom row: point-wise accuracy comparison comparing the mean absolute error of the coarse gridded satellite products (blue) to the mean absolute error of the GenLLC ensemble mean (red solid), and the CRPS of the GenLLC ensemble members (red hatched) for, from left to right, SSH, SST, SSS, and vorticity.}\label{fig:extended:osse_metrics}
\end{figure}

\begin{figure}[t]
\includegraphics[width=\textwidth]{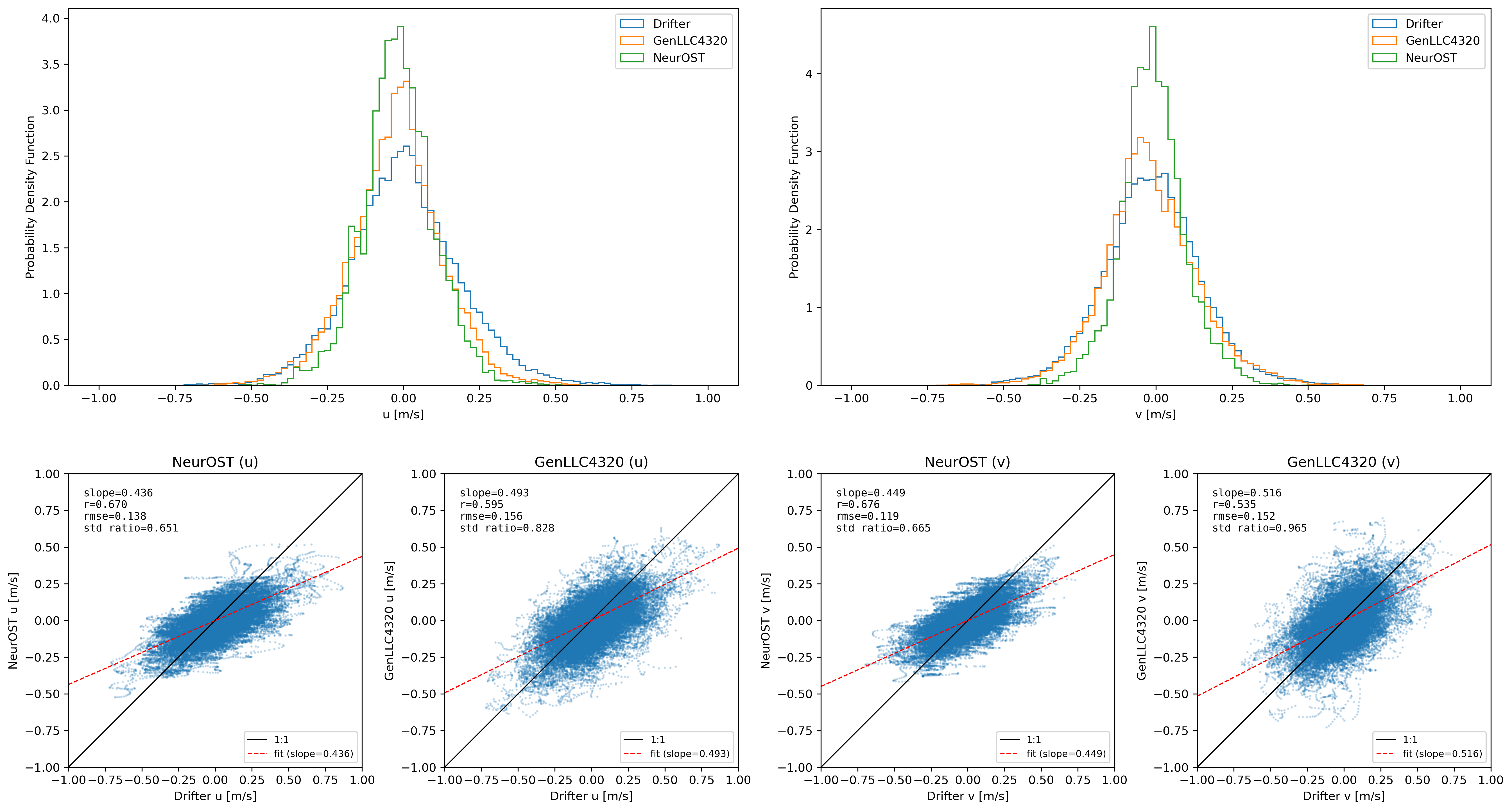}
\caption{Comparison of GenLLC surface currents to independent surface drifter observations from Global Drifter Program. PDF of (a) zonal and (b) meridional surface current velocities, comparing GenLLC (orange) and NeurOST (green) to drifter observations (blue). Scatter plots of predicted velocity versus drifter observations for (c,e) NeurOST and (d,f) GenLLC.}\label{fig:extended:drifter_validation}
\end{figure}

\begin{figure}[t]
\includegraphics[width=\textwidth]{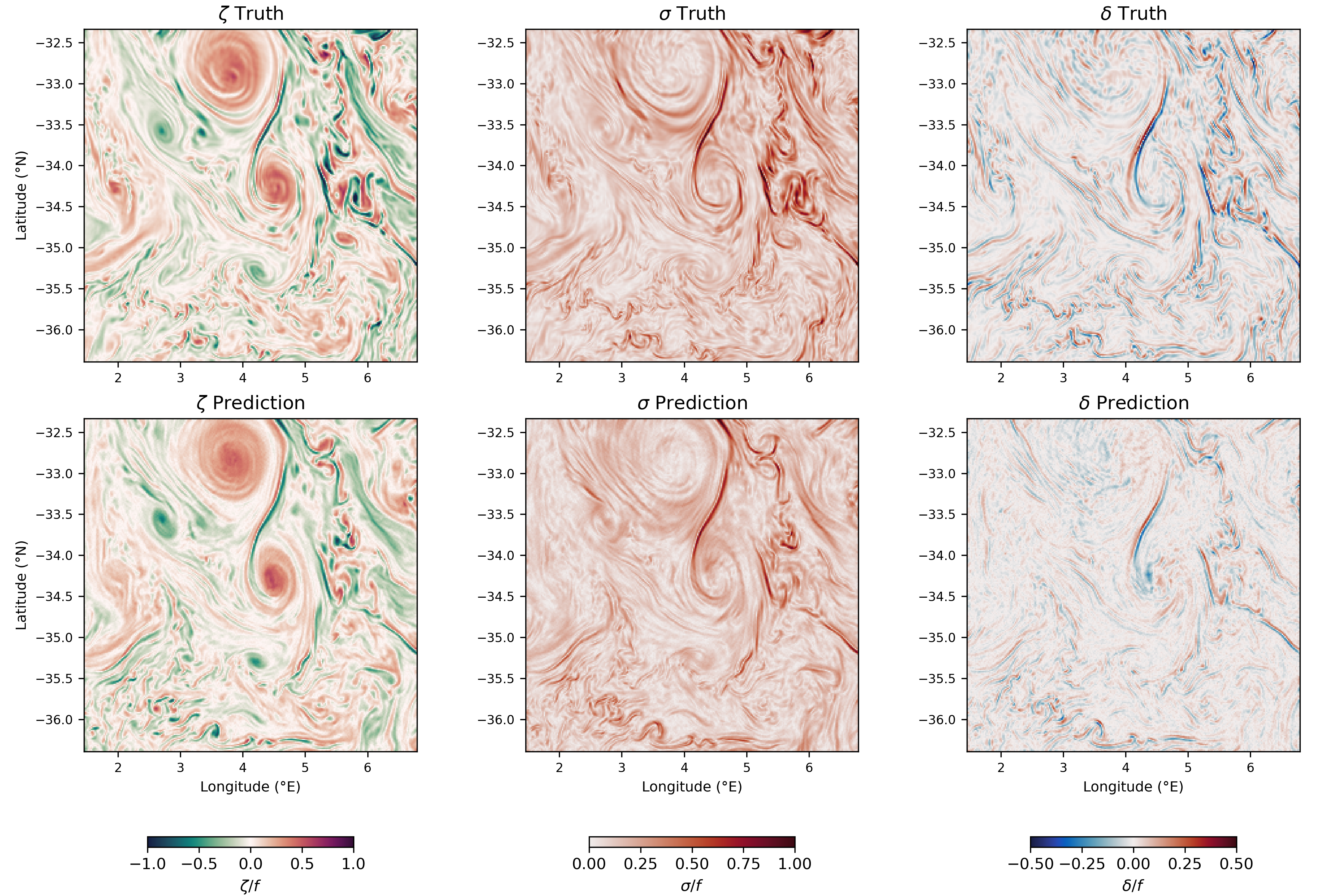}
\caption{Vorticity, strain, and divergence for a snapshot in the Agulhas Ring Path comparing LLC4320 ground truth (top row) to GenLLC prediction (bottom row).}\label{fig:extended:osse_vort_strain_ring_path}
\end{figure}

\begin{figure}[t]
    \begin{subfigure}{\textwidth}
        \includegraphics[width=\textwidth]{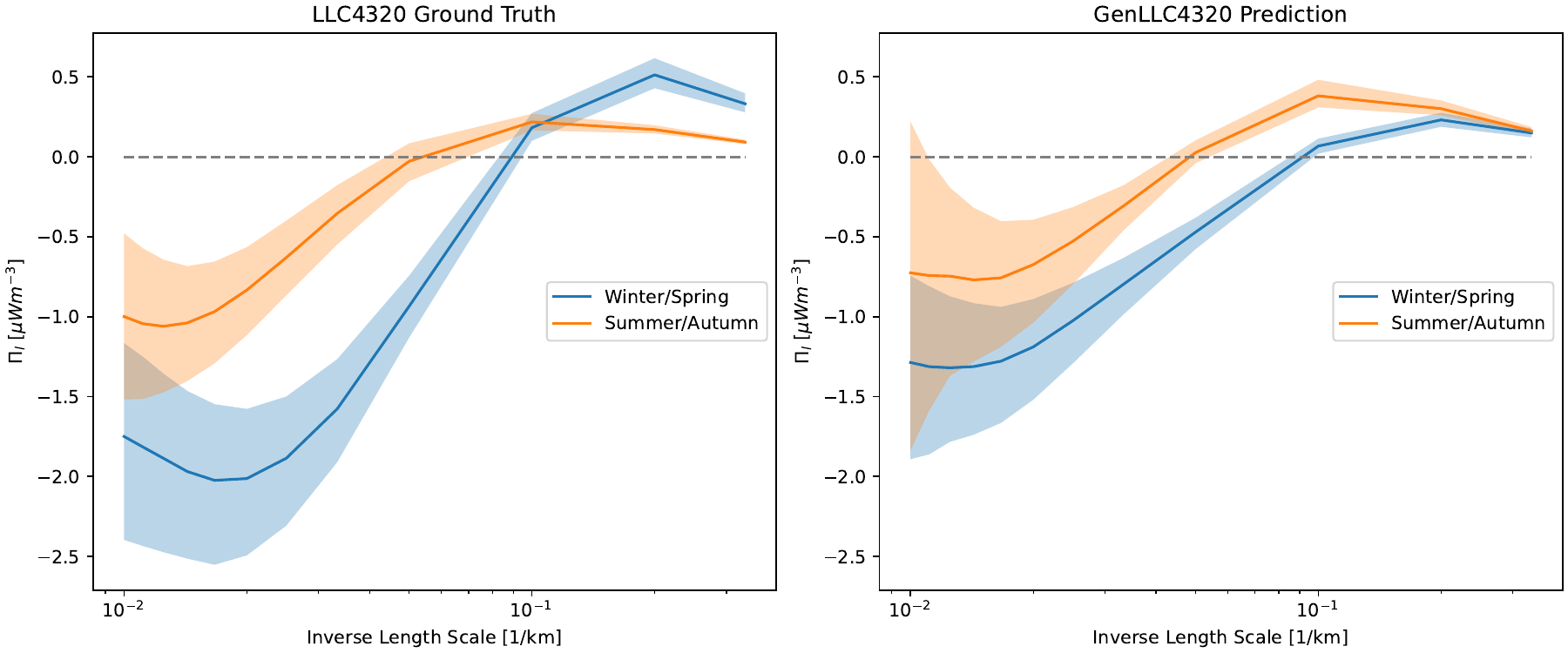}
        \caption{Ring Path}
    \end{subfigure}\\[1em]
    \begin{subfigure}{\textwidth}
        \includegraphics[width=\textwidth]{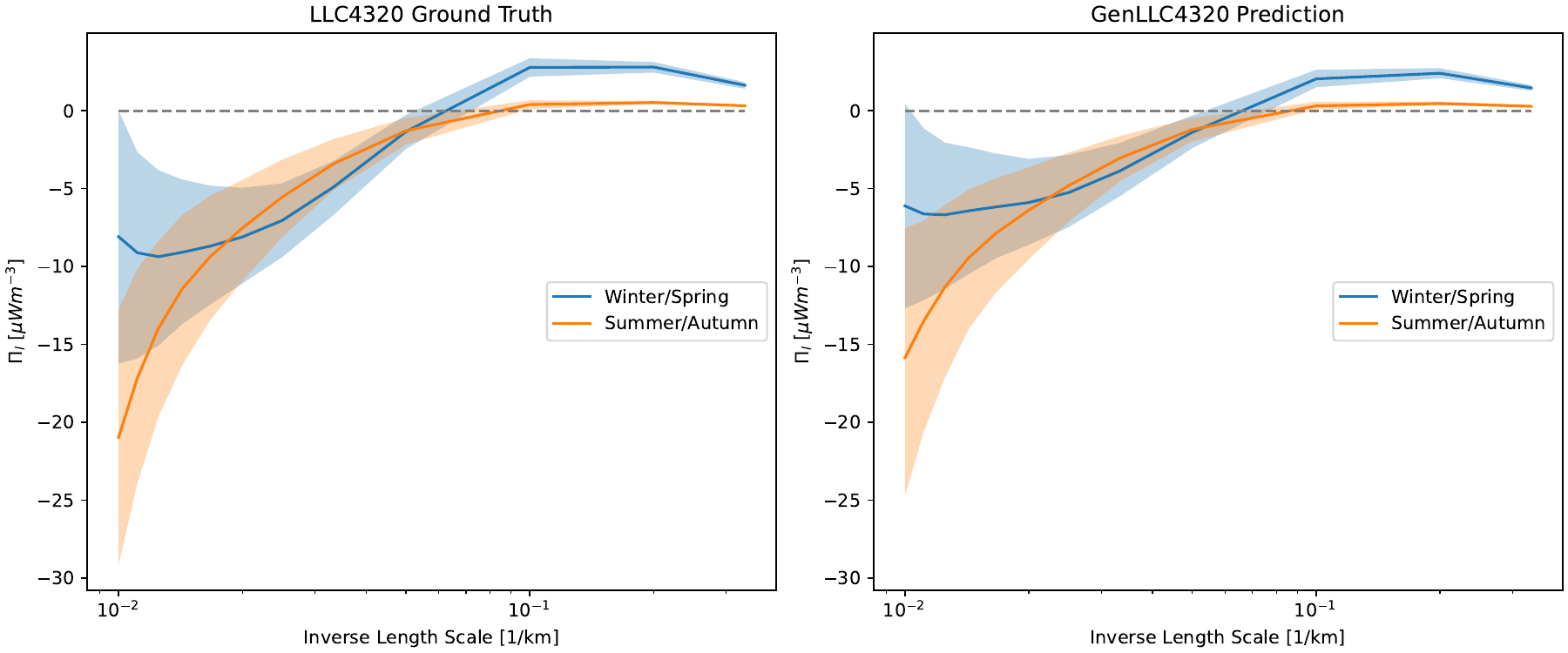}
        \caption{Return Current}
    \end{subfigure}
    \caption{KE cascade comparing GenLLC to LLC4320 ground truth in OSSE setting for the Agulhas Ring Path and Agulhas Return Current regions.}\label{fig:extended:osse_cascade}
\end{figure}

\begin{figure}[t]
\includegraphics[width=\textwidth]{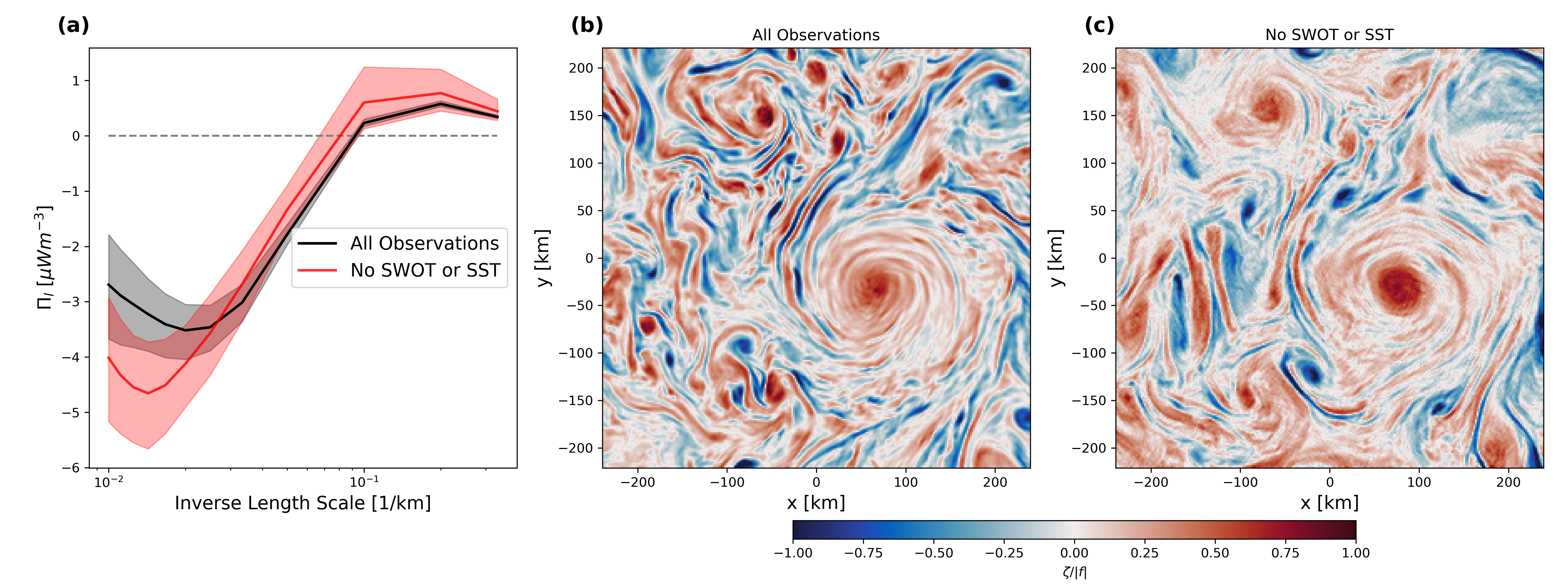}
\caption{Effect of high-resolution satellite observation constraints on the inferred kinetic energy cascade. (a) Winter/spring cross-scale kinetic energy flux from GenLLC with and without assimilation of SWOT SSH and high-resolution SST. Shading shows the 95\% confidence interval from bootstrapping across windows. (b) GenLLC vorticity snapshot with high-resolution observations assimilated, and (c) same as (b) but without assimilation of high-resolution observations.}\label{fig:extended:high_res_obs}
\end{figure}

\begin{figure}[t]
    \begin{subfigure}{\textwidth}
        \includegraphics[width=\textwidth]{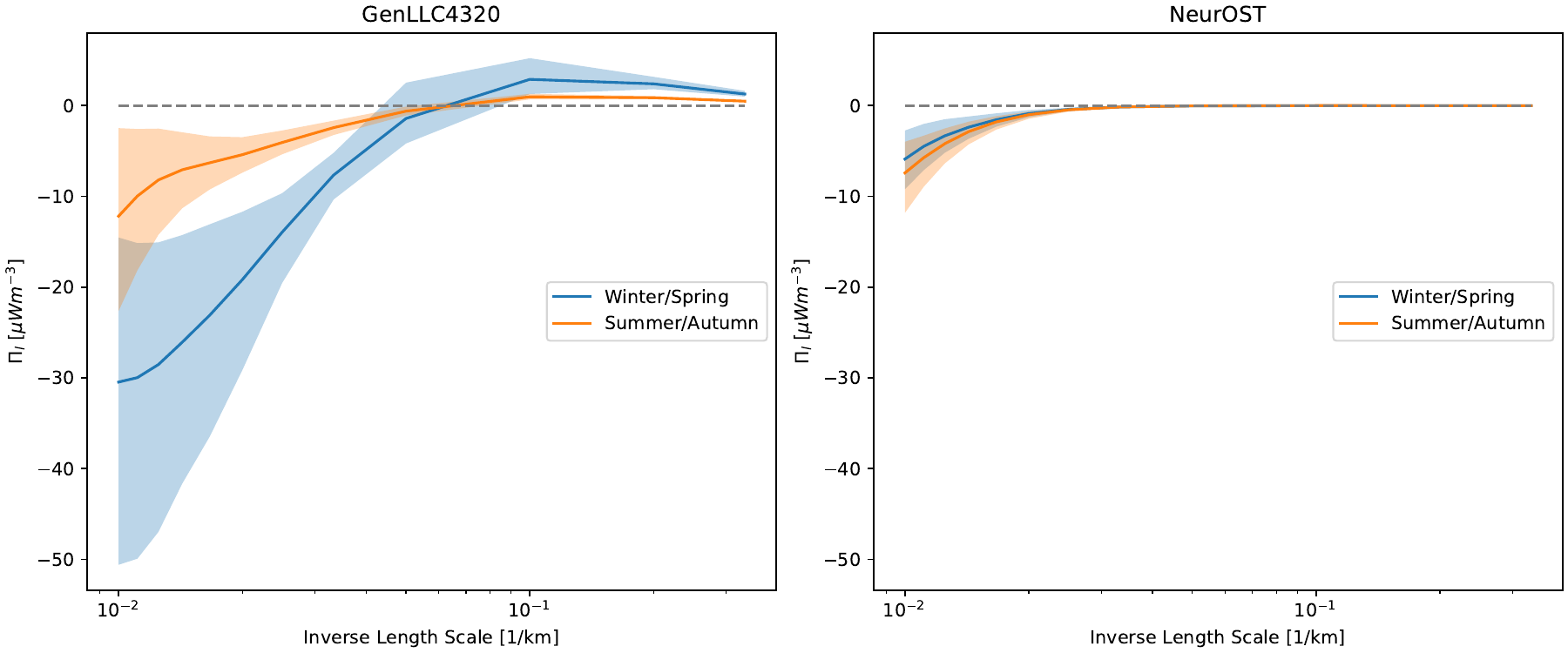}
        \caption{Real-World Observations }
    \end{subfigure}\\[1em]
    \begin{subfigure}{\textwidth}
        \includegraphics[width=\textwidth]{figs/osse_cascade_arc.pdf}
        \caption{OSSE}
    \end{subfigure}
    \caption{KE cascade in the energetic Agulhas Return Current region both in the real-world and OSSE settings.}\label{fig:extended:cascade_return_current}
\end{figure}

\begin{figure}[t]
\includegraphics[width=\textwidth]{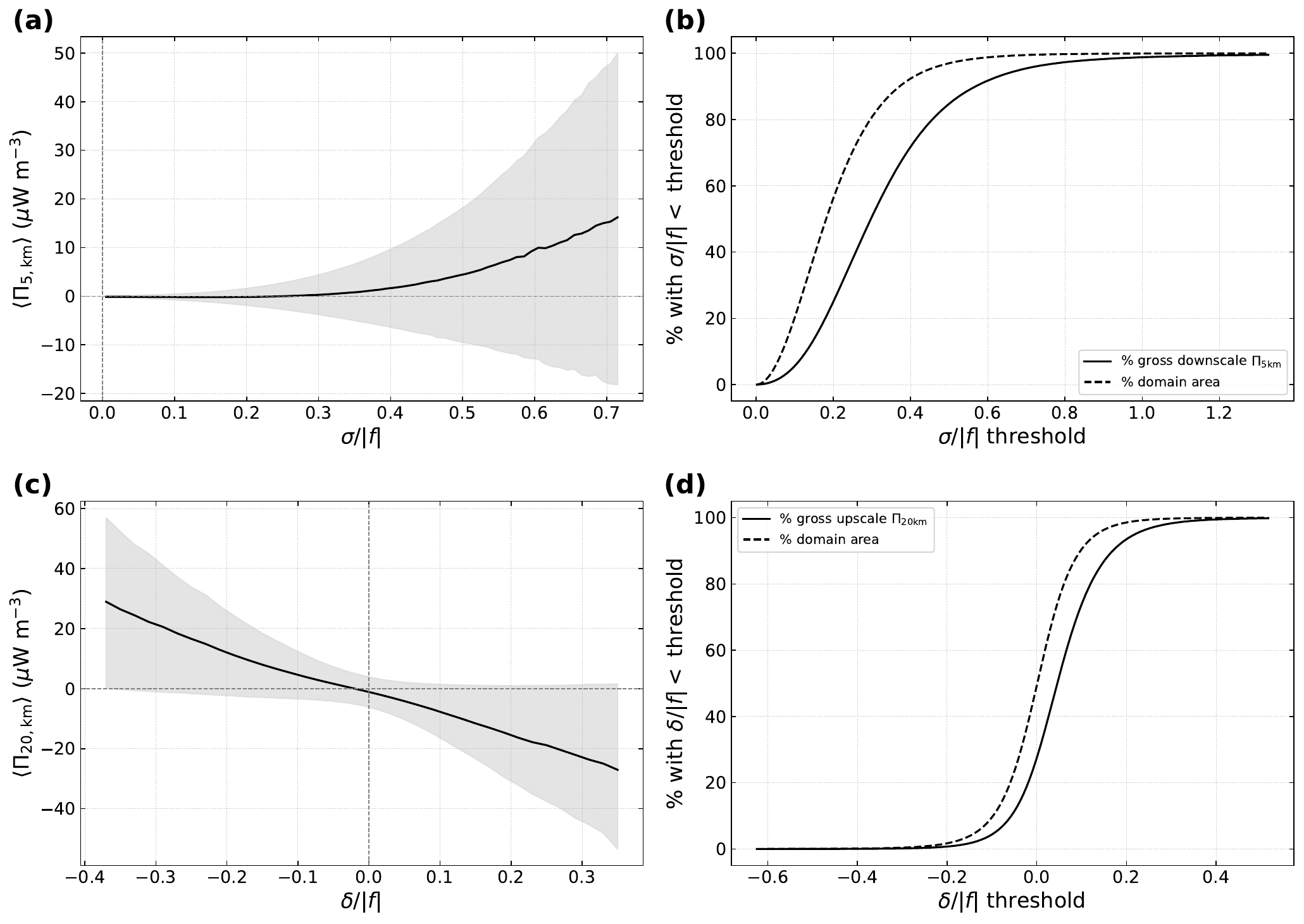}
\caption{Relationship between cross-scale fluxes and strain/divergence. (a) Conditional mean of the 20 km cross-scale KE flux, $\Pi_{5km}$, against strain, $\sigma$, with shading showing 16th and 84th percentile range. (b) Percentage coverage of domain area with $\sigma$ below threshold (dashed) compared to that areas percentage contribution to the domain-integrated gross downscale $\Pi_{5km}$. (c) Conditional mean of the 20 km cross-scale KE flux, $\Pi_{20km}$, against divergence, $\delta$, with shading showing 16th and 84th percentile range. (d) Percentage coverage of domain area with $\delta$ below threshold (dashed) compared to that areas percentage contribution to the domain-integrated gross upscale $\Pi_{20km}$.}\label{fig:extended:pi_v_div}
\end{figure}

\clearpage

\section*{Supplementary Information}

\renewcommand{\figurename}{Supplementary Fig.}
\setcounter{figure}{0}

\renewcommand{\tablename}{Supplementary Table}
\setcounter{table}{0}

\setcounter{section}{0}

\section{Additional evaluations of GenLLC surface state estimation}

\subsection{Extended discussion of state estimation validation results from main text}

Quantitative validation of the GenLLC state estimates is presented in Extended Data Figs.~\ref{fig:extended:real_full_snapshot}-\ref{fig:extended:drifter_validation}. Here we provide an expanded discussion of these validation results.

We evaluate the point-wise accuracy of the state estimates using the Continuous Ranked Probability Score \citepmeth[CRPS;][]{gneiting2007strictly,hersbach2000decomposition}, a scoring rule that jointly assesses the accuracy and calibration of probabilistic predictions by measuring the distance between the predicted ensemble distribution and the observed value. The CRPS decomposes into the mean absolute error of the ensemble mean minus a term rewarding ensemble spread that is proportional to forecast uncertainty, penalizing overconfident predictions where ensemble spread is low but errors are large. For a deterministic prediction, CRPS reduces to the mean absolute error, providing a unified framework for comparing GenLLC ensemble predictions against deterministic products such as coarse gridded satellite products.

\subsubsection{OSSE}

We first evaluate GenLLC in the OSSE context (Methods). To quantitatively evaluate reconstruction accuracy beyond a single snapshot, we evaluate reconstructions over 22 well-observed windows in the Return Current, generating a 20-member ensemble of state estimates for each. We compare the performance of GenLLC to the simulated coarse gridded satellite products (Extended Data Fig.~\ref{fig:extended:osse_metrics}). The spectra for the coarse satellite products for all variables strongly under-estimates small-scale variance when compared to the LLC4320 ground truth, while the GenLLC ensemble members each exhibit physically realistic spectra which closely track the ground truth down to the grid scale. The GenLLC ensemble mean is smoother at the smallest scales, reflecting the under-constrained nature of small-scale variability in gaps between observations. Deterministic methods would suppress these small-scale features entirely and regress to the ensemble mean. Across all variables, the GenLLC ensemble state estimates outperform the coarse gridded satellite products in point-wise accuracy, as shown by their lower CRPS values. For SSH and SST, the significantly lower CRPS values are in part due to the the assimilation of temporally adjacent observations, while SSS reflects the inherently low-resolution nature of satellite SSS observations. GenLLC exploits high-resolution SST to recover much of the submesoscale SSS signal. For vorticity, the GenLLC ensemble members have a higher mean absolute error than is obtained applying geostrophic balance to coarse gridded SSH, but lower CRPS confirms that GenLLC generates a well-calibrated probabilistic prediction whose ensemble spread appropriately reflects reconstruction uncertainty.

\subsubsection{Real World}

We validate GenLLC in the real world context by comparing satellite observable variables against withheld satellite observations (Methods). Direct validation of the reconstructed submesoscale vorticity is challenging given the absence of synoptic current observations, but applying geostrophic balance to an independent withheld SWOT pass reveals submesoscale eddy-like structures coinciding with the aggregation of submesoscale features generated by GenLLC in the center of the domain (Extended Data Fig.~\ref{fig:extended:real_full_snapshot}). The stronger mesoscale eddy signature in the northern part of the domain reconstructed by GenLLC relative to NeurOST is corroborated by the geostrophic vorticity from withheld SWOT SSH. 

Quantitatively, the spectra of the real world GenLLC state estimates show significantly stronger variability at small scales than gridded satellite products for SSH, SST, and SSS, and for surface currents when compared to geostrophic currents from NeurOST (Extended Data Fig.~\ref{fig:extended:real_metrics}). The CRPS and mean absolute error for SST are both significantly lower than that of the gridded satellite product when evaluated against withheld SEVIRI observations. In contrast, the SSH mean absolute error and CRPS appear marginally higher in GenLLC when compared against withheld SWOT observations than for NeurOST. This likely reflects the limited spatial coverage when a full SWOT pass is withheld, leaving substantial portions of the domain unobserved at any timestep which the diffusion prior does not fully overcome. Nonetheless, the reconstruction errors for SSH are only marginally higher than NeurOST. Since cascade diagnostics are computed using all available SWOT observations with none withheld for validation, the observational coverage in the cascade analysis is more spatially comprehensive, covering most of the domain in a 2.5 day window.

We additionally compare GenLLC surface velocities against \textit{in situ} observations from Global Drifter Program drifters. We focus on the Ring Path, which is more densely sampled than the Return Current, where energetic background currents advect drifters quickly out of the region. This comparison is complicated by what drifters measure: the total surface velocity, which includes not only the geostrophic and ageostrophic eddy turbulence GenLLC targets but also Ekman currents, near-inertial oscillations, and motions faster than the 12-hour GenLLC timestep. We low-pass filter the drifter data to remove sub-inertial frequencies, but their Lagrangian nature means they provide `ground truth' only in a loose sense.
A clear pattern nonetheless emerges. NeurOST systematically underestimates the drifter velocities, with a narrower velocity distribution and a predicted-versus-observed slope well below the ideal 1:1 line (Extended Data Fig.~\ref{fig:extended:drifter_validation}). GenLLC underestimates them too, but less systematically, capturing stronger velocity extremes and yielding a slope closer to 1:1. Viewed instead through point-wise metrics (RMSE, correlation), NeurOST has marginally lower errors. This trade-off matches \citetmeth{coadou2025resolving}, who found the same when validating SWOT geostrophic velocities against drifters: high-variance, submesoscale-resolving predictions tend to incur marginally higher point-wise errors than smooth, low-resolution ones.

\subsection{Testing the effect of super-inertial SSH observation contamination on GenLLC state estimates}

\begin{figure}[t]
\includegraphics[width=\textwidth]{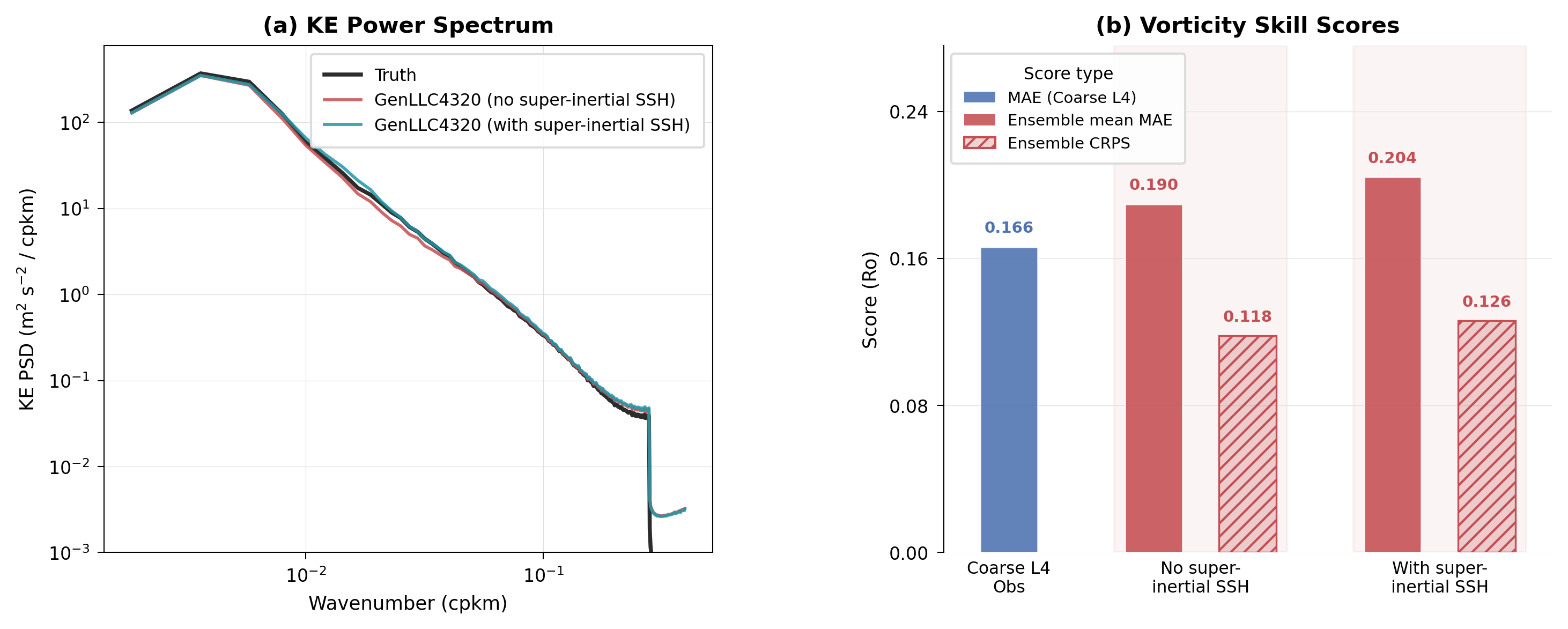}
\caption{Degradation of GenLLC performance in the Ring Path region when the assimilated synthetic SWOT SSH observations are contaminated with super-inertial signals such as internal tides for (a) KE spectra and (b) vorticity skill scores. In both cases, the ground truth is the LLC velocity field with the super-inertial signals filtered out and we compare GenLLC performance when the assimilated SWOT SSH does and does not contain super-inertial SSH contamination.}\label{fig:osse_subinertial_ablation}
\end{figure}

Real world SWOT SSH observations will contain a degree of signal contamination from sub-inertial internal gravity waves whereas our diffusion prior was trained to generate and assimilate SSH with sub-inertial internal tides filtered as a pre-processing step. To assess the potential effect of assimilating SSH observations with the super-inertial SSH signals contaminating the SSH we evaluate GenLLC using simulated satellite observations from LLC4320 but this time assimilate the unfiltered, total SSH, taking a randomized hourly time-step from within the prior's 12-hour step to reflect that SWOT observes instantaneous SSH rather than a 12-hour average (Supplemntary Figure \ref{fig:osse_subinertial_ablation}). The target variable for surface currents remains the sub-inertial filtered data, so that we can assess the bias introduced in the prediction by super-inertial signals in the SSH observations. The GenLLC state estimation metrics remain largely unchanged from those presented in the main text, with a minor degradation in vorticity CRPS the only apparent sign of degraded state estimation due to super-inertial SSH contamination, and with GenLLC still outperforming the coarse L4 observations in terms of CRPS even with contamination from super-inertial SSH. Looking at the reconstructed kinetic energy spectra, there is a modest signature of enhanced variance at scales between 100 and 10 km when super-inertial SSH signals are assimilated, suggesting that the diffusion prior is incorrectly translating some signatures of unbalanced internal tides in SSH (which were filtered out during training) into associated surface currents. While the effect of real world super-inertial SSH contamination on GenLLC state estimates remains an open question, and there is a clear need for novel methods to filter these signals from SWOT observations, the relatively minor state estimation degradation here increases confidence in the robustness of our observational constraints on the submesoscale cascade. Furthermore, LLC4320 has been found to have overly-energetic internal tides \citepmeth{savage2017spectral,yu2019surface,luecke2020statistical,arbic2022frequency}, so the effect in the real world could be smaller than that shown here.

\section{Additional Methods}

\subsection{Video diffusion prior architecture and training details}

GenLLC employs a UNet backbone neural network architecture with the augmentations widely used in the diffusion model literature \citepmeth{song2020score}: self attention near the bottleneck, noise level conditioning by performing a sinusoidal embedding of the scalar noise level to create a conditioning image tensor with multiple frequency channels, and replacing batch normalization with layer normalization to improve training stability when batches contain samples with very different noise levels. To process the time dimension in our architecture, we apply pixel-wise temporal attention at a number of resolution layers within the encoder-decoder path, folding time along the batch dimension for all other operations to process each timeframe independently using the same weights. We follow the Elucidated Diffusion Modeling (EDM) framework \citepmeth{karras2022elucidating} for training our prior, training a denoiser neural network with additive Gaussian noise applied at noise levels between $\sigma_{min}$ and $\sigma_{max}$. We follow the arguments of \citetmeth{brenowitz2025climate} when selecting these hyperparameters for geophysical data (see Supplementary Table \ref{tab:training_hypers}) and replace the log-normal noise distribution used in \citetmeth{karras2022elucidating} with the log-uniform distribution used in \citetmeth{brenowitz2025climate}. We additionally condition our diffusion prior on time of day, and day of year through additional sinusoidal embeddings \citepmeth{brenowitz2025climate} to provide context on the diurnal and seasonal cycle both of which impact submesoscale surface ocean dynamics. 

During model development we found that training a single diffusion backbone over the full range of noise levels, $\sigma$, led to unsatisfactory performance, underlining the challenge of training a single model to jointly handle a wide range of dynamics from mesoscales down to submesoscales. We thus instead pursued a `mixture of experts' approach, where separate models were trained for low and high noise levels respectively, allowing each to focus on a distinct scale range, since noise levels in diffusion models are loosely equivalent to spatial scales with higher noise levels corresponding to larger scales. This strategy is similar to that employed to prevent overfitting at large scales in \citetmeth{brenowitz2025climate} except here we train separate models from scratch on the different noise ranges rather than training a single model and picking different checkpoints for different noise ranges. The small and large $\sigma$ experts are trained with overlapping $\sigma$ ranges, and at inference we switch which model is called at a prescribed handover $\sigma$ at the center of the overlap range. During training, we also conditioned the model on patch-averaged surface wind stress, surface heat flux, and surface freshwater flux, however at inference we found these fluxes to be largely ignored by the model and all results presented in the main text used placeholder values of zero for these fluxes for simplicity of implementation.

\begin{table}[h]
\centering
\begin{tabular}{lc}
\toprule
Hyperparameter & Value \\
\midrule
$\sigma_{min}$ & $10^{-3}$ \\
$\sigma_{max}$ & 200 \\
Resolution levels for self attention  & 32, 16 \\
Resolution levels for temporal attention  & 64, 32, 16 \\
Number of channels in first UNet block  & 128 \\
Channel multipliers per resolution  & 1, 2, 2, 3, 4 \\
UNet blocks per resolution level & 3 \\
Learning rate & $10^{-4}$ \\
Batch size & 48 \\
Training resources & 3 days on 8 A100 GPUs for each of the 2 experts \\
Small $\sigma$ expert noise range & $[10^{-3}, 1]$ \\
Large $\sigma$ expert noise range & $[10^{-1}, 200]$ \\
Handover $\sigma$ between experts & 0.5 \\
\bottomrule
\end{tabular}
\caption{Training and architecture hyperparameters.}
\label{tab:training_hypers}
\end{table}

\subsection{SDA hyperparameters}

At inference, we use the score-based data assimilation (SDA) framework \citepmeth{rozet2023score} to guide the generation of the diffusion prior with sparse observations. Since the SDA algorithm was developed for the variance-preserving stochastic differential equation (VPSDE) sampler, as opposed to the probability flow sampler used in EDM \citepmeth{karras2022elucidating}, we translate our EDM diffusion prior to the VPSDE formulation using the formula derived in the appendix of \citetmeth{manshausen2024generative}. Supplementary Table \ref{tab:inference_hypers} details the values used in this study for the SDA hyperparameters which are described in \citetmeth{rozet2023score}. These values were arrived at through explorations carried out on simulated observations from the training partition of the LLC4320 data. We found the fidelity of reconstructions to improve as $\Gamma$ was reduced, but with increasing instances of numerical instability below the value used here, and even with this value there were a small minority of ensemble members that went numerically unstable at inference and we prune before the analyses presented in the main text since they are easily identifiable by their pixelated, un-physical predictions. We found applying Langevin Monte Carlo corrector steps only at low noise levels beneath $t=0.5$ helped to stabilize sampling while retaining the benefits of corrector steps; with no corrector steps the generated states tended to smooth out variance at submesoscales. 

\begin{table}[h]
\centering
\begin{tabular}{lc}
\toprule
Hyperparameter & Value \\
\midrule
$\Gamma$ & 0.05 \\
$\eta$ & $10^{-4}$ \\
Number of steps & 196 \\
Number of corrector steps above $t=0.5$ & 0 \\
Number of corrector steps below $t=0.5$ & 5 \\
$\tau$ & 0.2 \\
\bottomrule
\end{tabular}
\caption{SDA inference hyperparameters.}
\label{tab:inference_hypers}
\end{table}

\subsection{Details of the satellite data products used in this study}

We use real satellite observations from the year 2024 in this study, and in the observing system simulation experiment take the sampling masks from this year of observations and apply it to the simulation.

The sparse, high-resolution SST observations are taken from the SEVIRI instrument onboard the Meteosat Second Generation (MSG) satellites. These satellites are on a geostationary orbit, providing continuous in time observations, albeit with coverage of only a limited geographical domain. We use the Level 3 collated product produced by OSI SAF which aggregates the observations onto a $0.05^{\circ}$ grid with hourly resolution, applies numerous atmospheric corrections, applies a cloud mask, and provides quality flags for each data point. Since our diffusion prior is trained to produce 12-hourly means, we composite the SEVIRI observations within a 12-hour window, discarding measurements not flagged "good" or "very good" and discarding grid points with fewer than 3 observations within the 12-hour window. We note that the cloud masking algorithm used in the production of this dataset appears to sometimes artificially mask out some oceanographic features like mesoscale eddies and submesoscale fronts which could in future be alleviated by using lower-level data products as was done in \citetmeth{lenain2025unprecedented}, though we prefer here to retain the atmospheric and sensor corrections made at Level 3.

The sparse, high-resolution SSH observations are taken from both conventional nadir altimeters and SWOT. The nadir altimeter data is taken from the CMEMS Level 3 product where numerous tidal and atmospheric corrections have been made as well as calibration between the different altimeter missions. For SWOT, we take wide-swath observations from the AVISO Level 3 2.5 km Expert product (v2.0) where tidal corrections, inter-calibration with conventional altimeters, atmospheric corrections, orbit roll correction, and quality flagging have all been implemented. Note, we use here the `unfiltered' SSH anomaly variable, preferring not to rely upon the experimental convolutional neural network denoising method implemented in the product which could obscure real signals of submesoscale dynamics. We use the 2.5 km product rather than the 250 m product since the grid resolution of LLC4320, and hence our diffusion prior, is considerably larger at 1.5-2 km and this mitigates potential contamination from surface waves.

In addition, we leverage coarse gridded satellite products for SSH, SST, and SSS both as a large-scale constraint in the assimilation and to provide a comparison to GenLLC state estimates. For SSH, we use the NeurOST product which is produced by synthesizing along-track nadir altimeter observations with gridded SST from the NASA MUR product using a neural network trained on past observations. The grid resolution is $1/10^{\circ}$, though the effective resolution is 90-120 km in wavelength in the region considered here. We also use surface geostrophic currents from the NeurOST product which have been validated against surface drifters \citepmeth{martin2024deep}, though we coarse-grain these with a 100 km Gaussian filter before assimilation to ensure we only assimilate geostrophic currents at larger scales where geostrophy is valid and where the NeurOST estimates will be most reliable. For SST, we use the REMSS MW OI gridded product which uses optimal interpolation to merge measurements from a number of missions equipped with microwave SST sensors into a $1/4^{\circ}$ product. While microwave observations are lower resolution than infrared, they can penetrate clouds, meaning this product appears to have a temporally consistent effective spatial resolution which is important for this study since we assimilate the product as a large-scale constraint. For SSS, we use the CMEMS Multi Observation Global Ocean Sea Surface Salinity and Sea Surface Density Level 4 product which uses a multivariate optimal interpolation algorithm combining satellite and in situ salinity observations with additional constraints from satellite SST to produce a $1/8^{\circ}$ product. When assimilating these coarse gridded satellite products in the generative data assimilation process we first coarse-grain the state estimate using prescribed coarse-graining scales chosen to approximate the effective resolution of the above products. These scales are detailed in Supplementary Table \ref{tab:sigmas}.

\begin{table}[h]
\centering
\begin{tabular}{lc}
\toprule
Hyperparameter & Value \\
\midrule
SSH & 30 km \\
SST & 15 km \\
SSS & 30 km \\
u/v & 100 km \\
$\sigma_{time}$ for all variables & 2.5 days \\
\bottomrule
\end{tabular}
\caption{Coarse-graining scales (Gaussian filter width, $\sigma$) applied to each variable before assimilating coarse gridded satellite products.}
\label{tab:sigmas}
\end{table}

\section{Validity of estimating the submesoscale kinetic energy cascade in small spatio-temporal windows}\label{app:small_patch_ablation}

\begin{figure}[t]
    \begin{subfigure}{\textwidth}
        \includegraphics[width=\textwidth]{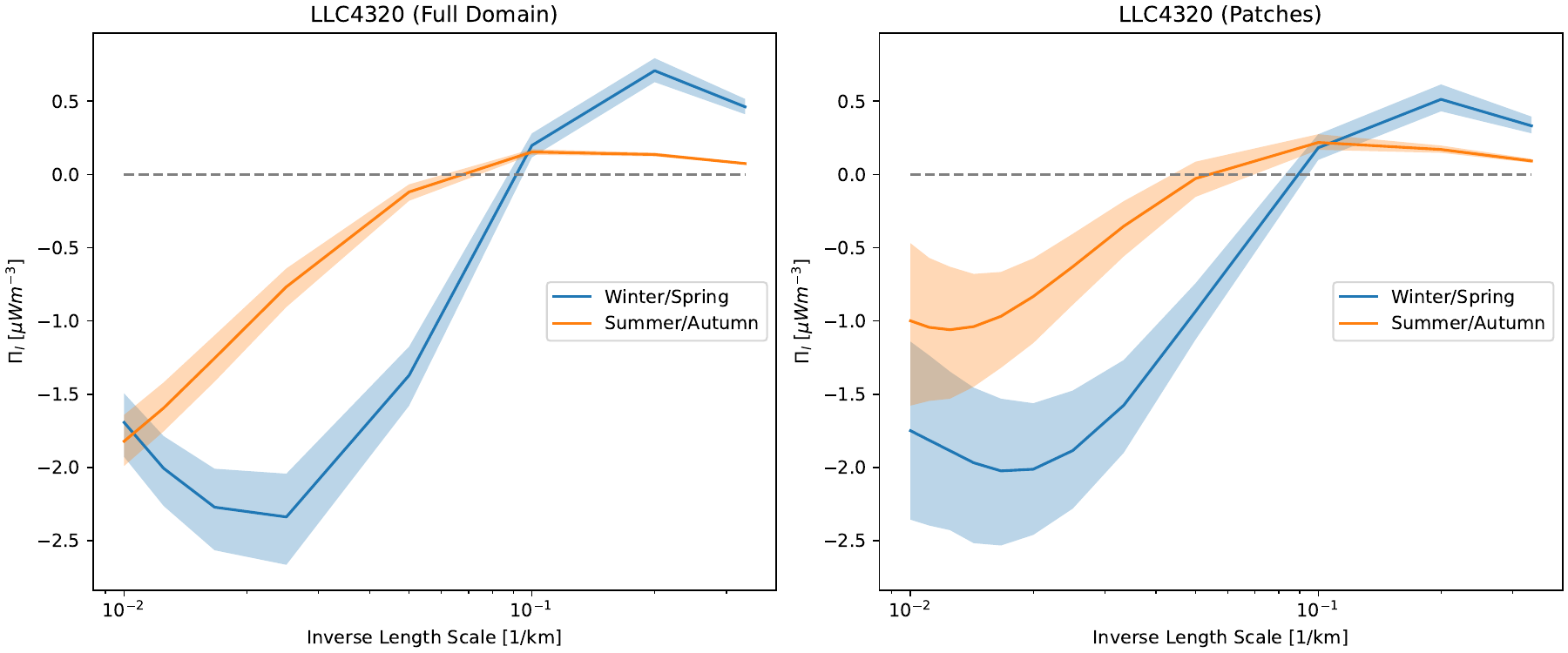}
        \caption{Ring Path}
    \end{subfigure}\\[1em]
    \begin{subfigure}{\textwidth}
        \includegraphics[width=\textwidth]{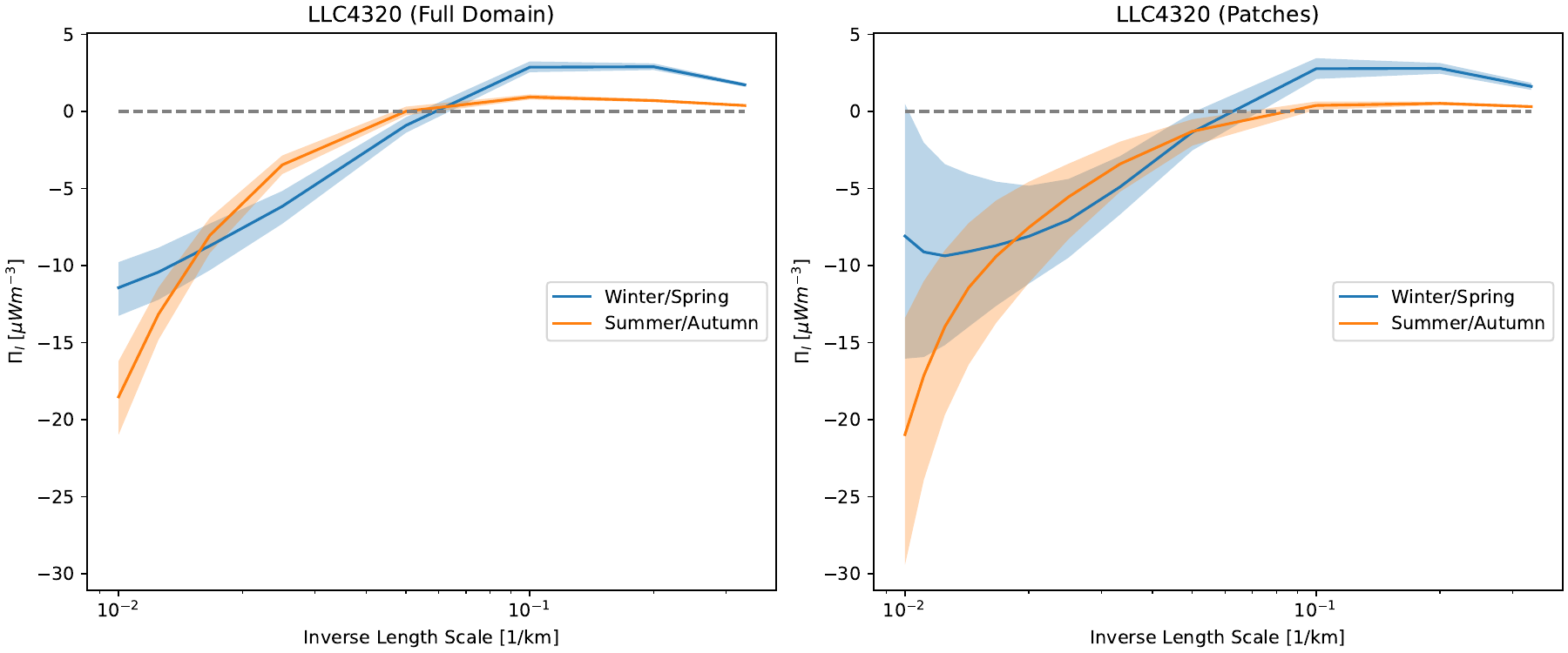}
        \caption{Agulhas Return Current}
    \end{subfigure}
    \caption{Comparison of the kinetic energy cascade for the LLC4320 ground truth inferred from the small well-observed windows used in the main text compared to computation on the full fields for the study region showing the validity of estimating KE cascade in small, well-observed patches.}\label{fig:osse_cascade_small_patch_ablation}
\end{figure}

Since our diffusion prior is restricted to generating a relatively restricted domain of 256x256 LLC4320 grid points ($\sim 400$~km) and relies upon good observational coverage both from SWOT SSH and SEVIRI SST, in the main text we diagnose the strength and seasonality of the kinetic energy cascade by applying the coarse-graining framework \citepmeth{aluie2018coarsegrain,storer2023flowsieve} within a set of well-observed windows and averaging over many such windows to get seasonal composite of the cascade in our larger study regions (the Return Current and the Ring Path). This limited sampling introduces an extra potential source of noise and error in the cascade estimates which we assess here. We compute the seasonal composites of the cascade from the LLC4320 ground truth from the full underlying fields on all time-steps in our study regions (which are $\sim1000$~km in size) and compare the result to that estimated from the small well-observed windows used in the main text (Supplementary Figure \ref{fig:osse_cascade_small_patch_ablation}). This comparison shows that in both regions the computation on the full domain results in the same qualitative conclusions about the cascade below 100 km, with the estimations in good agreement across the scale range considered in this study. The noise and bias from our limited sampling becomes more pronounced at larger scales approaching 100 km, which is reflected in a corresponding broadening of the bootstrapped confidence intervals. The increased uncertainty at larger scales stems from excluding a border region with width that of the coarse-graining filter scale before computing the domain-averaged cross-scale flux to avoid contamination from edge effects. The close correspondence between the small patch and full domain computations increases confidence that the results presented in the main text are not influenced of sampling artifacts and provide a robust estimate of the kinetic energy cascade below 100 km.

\bibliographymeth{sn-bibliography}

\end{document}